\documentclass[twocolumn,
aps,
prx,
showpacs,
amsmath,
amssymb,
floatfix,
longbibliography,
superscriptaddress,
eprint]
{revtex4-1}

\usepackage{subfigure}
\usepackage{amsfonts} 

\usepackage{xcolor}
\definecolor{darkblue}{RGB}{0,0,150}
\definecolor{nightblue}{RGB}{0,0,100}

\usepackage{graphicx,mathtools,bm,bbm}
\usepackage{MnSymbol}
\usepackage[
colorlinks,
citecolor=darkblue,
linkcolor=darkblue,
urlcolor=nightblue]{hyperref}

\usepackage[english]{babel}
\usepackage[babel,kerning=true,spacing=true]{microtype}
\usepackage[utf8]{inputenc}

\usepackage{soul}

\renewcommand{\vec}[1]{\bm{#1}}

\usepackage{amssymb}
\usepackage{pifont}

\begin{document}

\title{Topological excitations at time vortices in periodically driven systems}
    
\author{Gilad Kishony}\email{gilad.kishony@weizmann.ac.il}
\affiliation{Department of Condensed Matter Physics, Weizmann Institute of Science, Rehovot, 76100, Israel}
\author{Ori Grossman}
\affiliation{Department of Condensed Matter Physics, Weizmann Institute of Science, Rehovot, 76100, Israel}
\author{Netanel Lindner}
\affiliation{Department of Physics, Technion, Haifa, Israel}
\author{Mark Rudner}
\affiliation{Department of Physics, University of Washington, Seattle, USA}
\author{Erez Berg}
\affiliation{Department of Condensed Matter Physics, Weizmann Institute of Science, Rehovot, 76100, Israel}

\begin{abstract}
We consider two-dimensional periodically driven systems of fermions with particle-hole symmetry. Such systems support non-trivial topological phases, including ones that cannot be realized in equilibrium. We show that
a space-time defect in the driving Hamiltonian, dubbed a ``time vortex,''  
can bind $\pi$ Majorana modes. A time vortex is a point in space around which the phase lag of the Hamiltonian changes by a multiple of $2\pi$. We demonstrate this behavior on a periodically driven version of Kitaev's honeycomb spin model, where $\mathbb{Z}_2$ fluxes and time vortices can realize any combination of $0$ and $\pi$ Majorana modes. We show that a time vortex can be created using Clifford gates, simplifying its realization in near-term quantum simulators. 
      
\end{abstract}

\maketitle

\section{Introduction}

Periodically driven quantum many-body systems are known to feature a richer topological structure than their non-driven counterparts \cite{Kitagawa2010,Jiang2011,Rudner2013,Nathan2015,Titum2016,Keyserlingk2016,Roy2016,Khemani2016,Else2016,Roy2017,oka2019floquet,rudner2020floquet}. This additional structure arises from the periodicity of the quasi-energy spectrum characterizing the Floquet states. For example, periodically driven systems support “anomalous” topological phases characterized by the presence of chiral edge states, despite all the Floquet bulk bands being topologically trivial~\cite{Rudner2013}. 
Systems with particle-hole symmetry (such as Bogoliubov excitations in a superconductor) can support topologically protected Majorana excitations whose quasi-energy is either zero or $\pi/T$ (known as 0 and $\pi$ Majorana modes, respectively), where $T$ is the driving period~\cite{Jiang2011,kitagawa2012observation}.
These Majorana modes may appear at defects, such as domain walls, edges, or vortices in a superconductor. 
Systems combining $0$ and $\pi$ Majoranas open new possibilities for quantum information processing~\cite{Liu2013,Bomantara2018,Bauer2019,Matthies2022}, not available in static systems.
The emergence of $\pi$ Majorana modes is closely related to discrete time crystallinity~\cite{Wilczek2012,Khemani2016a,Else2016a}. 

\begin{figure}
\centering\includegraphics[ width=1\columnwidth ]{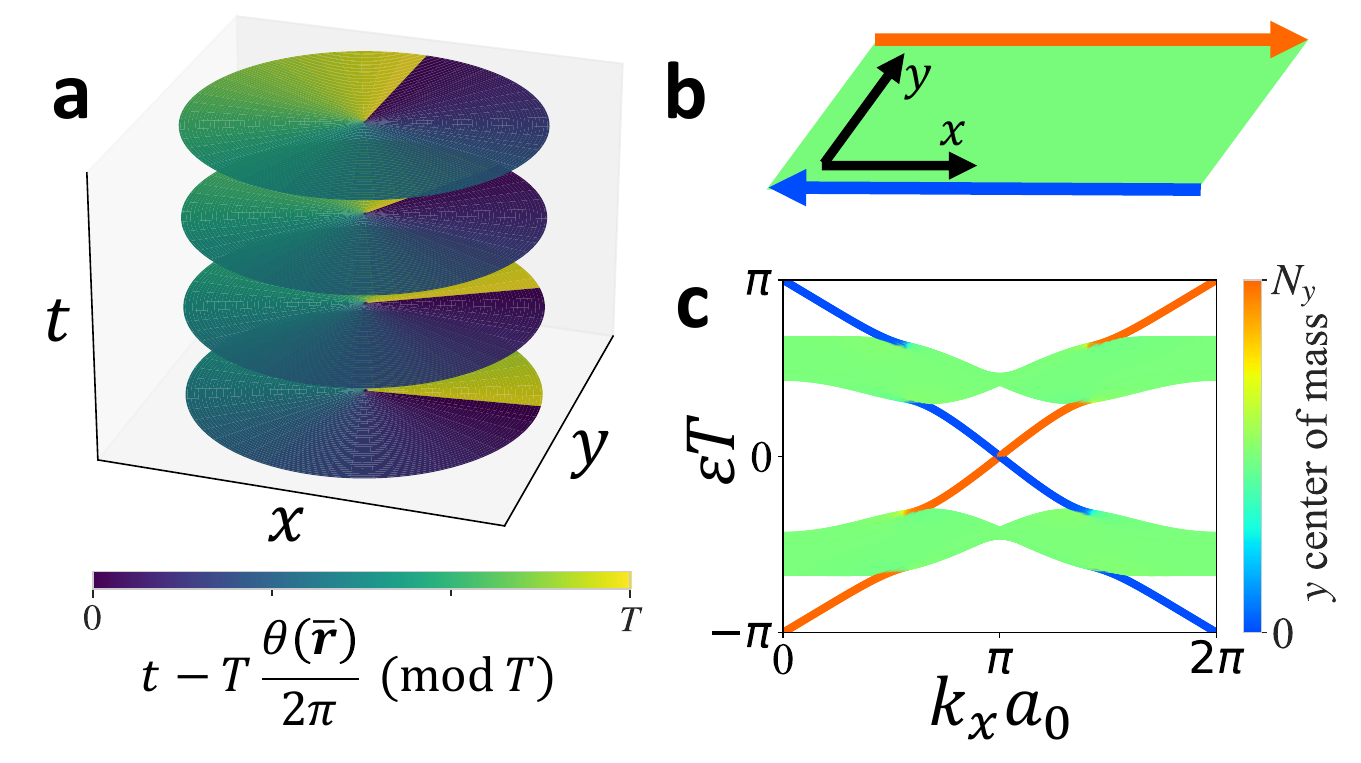}
\caption{\textbf{The time vortex defect and the driven Kitaev honeycomb model.} \textbf{(a)} An illustration of the time vortex defect in space-time. \textbf{(b)} 
The driven Kitaev honeycomb model in the anomalous phase with periodic boundary conditions on the $x$ axis and open boundary conditions on the $y$ axis has chiral edge states. \textbf{(c)}
The corresponding quasi-energy spectrum at $\Delta t=0.5 T$, $\mathcal{J}^a_0\Delta t=0.9\pi/4$.
}
\label{fig: space-time vortex and spectrum on cylinder}
\end{figure}

Periodically driven systems may possess 
space-time symmetries that refine the 
classification of topological phases~\cite{Morimoto2017,Peng2019,Peng2020,Peng2021,Peng2022}, as well as novel kinds of space-time topological defects~\cite{Katan2013,Yao2017}.
In two-dimensional periodically driven systems, it is natural to define a space-time topological defect, a ``time vortex,'' which is our focus in this work. 
A time vortex is a point in space around which the driving Hamiltonian's phase lag 
changes as a function of position, returning to itself modulo an integer multiple of $2\pi$ 
upon 
encircling the vortex (Fig.~\ref{fig: space-time vortex and spectrum on cylinder}a). 
We find that a time vortex can bind topological excitations in a system of fermions with particle-hole symmetry and no time reversal symmetry (symmetry class D~\cite{Altland1997}). Such a system may host chiral Majorana edge modes crossing the gaps at quasi-energy 0 and $\pi/T$~\cite{Jiang2011}. We show that a time vortex binds a Majorana $\pi$ mode at its core if the number of chiral edge states through quasi-energy $\pi/T$ is odd.

As a possible realization of this phenomenon, we consider a periodically driven version~\cite{Po2017,Fulga2019} of Kitaev's spin model on the honeycomb lattice~\cite{Kitaev2006}. Time reversal symmetry is broken by the chirality of the driving protocol \footnote{By time reversal symmetry, we mean that there is an anti-unitary operator $\mathcal{T}$ such that the time-dependent Hamiltonian satisfies $H(t)=\mathcal{T} H(T-t) \mathcal{T}^{-1}$, for an appropriate choice of the origin of time.}. As in Kitaev's original model, the system can be described in terms of emergent fermions with particle-hole symmetry, coupled to a static background $\mathbb{Z}_2$ gauge field. For a certain parameter range, the model realizes an anomalous Floquet phase with chiral Majorana edge states traversing the gaps around 
quasienergy 0 and $\pi/T$; 
in this phase, a $\mathbb{Z}_2$ flux carries both  Majorana 0 and $\pi$ modes. We demonstrate explicitly that a time vortex in this system binds a $\pi$ Majorana mode, but not a $0$ mode. Remarkably, this implies that by combining $\mathbb{Z}_2$ fluxes and time vortices, one can realize any combination of localized $0$ and $\pi$ Majorana modes.
The anomalous phase with or without a time vortex can be realized simply by periodically applying a sequence of Clifford gates to a 2D array of qubits, making it particularly suitable for implementation on near-term quantum hardware. 

\section{Results}
\subsection{The Floquet time vortex}
\label{sec: time vortex - general}
To define a time vortex, we consider a reference time-periodic tight-binding Hamiltonian $H_0$, with period $T$, defined on a 2D lattice: 
\begin{align}
H_0(\bm{r},\bm{r}',t+T) = H_0(\bm{r},\bm{r}',t),
\end{align}
where $\bm{r}$, $\bm{r}'$ denote the  sites of the system. 
To introduce a time vortex into $H_0$, we define a modified Hamiltonian 
$H_{\rm{v}}$
with a spatially dependent time delay: 
\begin{align}
H_{\rm{v}}(\bm{r},\bm{r}',t;\theta) = H_0\left[\bm{r},\bm{r}',t-\frac{T}{2\pi}\theta(\bar{\bm{r}})\right], \label{eq:H_theta}
\end{align}
where $\bar{\bm{r}} = \frac12(\bm{r}+\bm{r}')$.
To describe a time vortex at the origin, we take $\theta(\bar{\bm{r}})$ to be given by
\begin{align}
\label{eq:theta} \theta(\bar{\bm{r}}) = 
\arg(\bar{x}+i\bar{y}),
\end{align}
where we measure positive angles in the counterclockwise direction from the vector $\hat{\bm{x}}$. 
Thus, compared to $H_{0}$, $H_{\rm{v}}$ has a position-dependent delay in its time dependence,
which smoothly accumulates a full period $T$ when going in a closed loop around the center of the time vortex. 
This configuration realizes a topological defect that cannot be smoothly deformed to a trivial spatial dependence. Fig.~\ref{fig: space-time vortex and spectrum on cylinder}(a) illustrates this space-time dependence.

Being a topological defect, it is natural to ask if the time vortex can carry topologically protected modes in its core. Here, we focus on systems in symmetry class D, such that the time-dependent Hamiltonian $H_0(t)$ itself is particle-hole symmetric {at each time $t$}, satisfying $\mathcal{P}^{-1} H_0(t) \mathcal{P}= -H_0(t)$ and $\mathcal{P}^2=1$, where $\mathcal{P}$ is an anti-unitary operator. It follows that $H_{\rm{v}}(t)$ is also in class D.
A Floquet system
in class D is characterized by 
a set of integer-valued topological invariants $\mathcal{W}_{\varepsilon}$ 
defined for each gap in the quasi-energy spectrum ($\varepsilon$ denotes a quasi-energy within a gap).
The invariant $\mathcal{W}_{\varepsilon}$ is equal to the net number of chiral edge states that appear within the corresponding gap, weighted by their chirality~\cite{Rudner2013} (see Fig.~\ref{fig: space-time vortex and spectrum on cylinder}b,c). Our main result relates the properties of a time vortex to the bulk invariants of the Floquet phase: we show that if $\mathcal{W}_{\pi/T}$ is an odd integer, then a time vortex carries a topologically protected Majorana mode with quasi-energy $\pi/T$. 


\subsection{Semiclassical analysis of the $\pi$ mode on the edge}
\label{sec: semiclassical}

We now formulate a semiclassical argument showing that inserting a time vortex through 
a system in a cylindrical geometry can toggle the existence of 
$\pi$ modes at both edges on and off. To this end, we consider a strip of size $L_x\times L_y$ with periodic boundary conditions along the $x$ direction 
[see Fig.~\ref{fig: space-time vortex and spectrum on cylinder}(b)]. 

We start with a tight binding Floquet model of fermions in class D, with translation invariance in the $x$ direction (i.e., {\it without} a time vortex): 
\begin{align}
H_0(\bm{r}+a_0\hat{\bm{x}},\bm{r}'+a_0\hat{\bm{x}},t)=H_0(\bm{r},\bm{r}',t),
\label{eq:Hrrp}
\end{align}
where $a_0$ is the lattice spacing. 
We assume that the Hamiltonian has a finite range $\ell$, such that the matrix element \eqref{eq:Hrrp} vanishes for $|\bm{r}-\bm{r}'|>\ell$.

{For constructing the Floquet states in the presence of a time vortex, it will be useful to consider the Floquet states of this translation-invariant problem with twisted boundary conditions parameterized by a phase $\alpha$.
In accordance with Bloch's theorem, we label the Floquet states of the system by a crystal momentum component $k$ along $\hat{\vec{x}}$ (around the cylinder) and band index $n$, along with the twist angle parameter $\alpha$:
\begin{equation}
    \label{eq:BlochThm} \psi_{nk}(\bm{r}+a_0\hat{\bm{x}},t;\alpha)=e^{ika_0}\psi_{nk}(\bm{r},t;\alpha),
\end{equation}
with the twisted boundary conditions
\begin{equation}
\label{eq:BCs} \psi_{nk}(\bm{r}+L_x\hat{\bm{x}},t;\alpha) =e^{i\alpha}\psi_{nk}(\bm{r},t;\alpha).
\end{equation}
The corresponding quasi-energies $\varepsilon_{nk}(\alpha)$ follow from application of Floquet's theorem,}
\begin{equation}
    \label{eq:FloquetThm} \psi_{nk}(\bm{r},t+T;\alpha) =e^{-i\varepsilon_{nk}(\alpha)T}\psi_{nk}(\bm{r},t;\alpha).
\end{equation}
{The twisted boundary conditions yield a discrete set of allowed crystal momentum values
\begin{align}
\label{eq:k_quant} k=\frac{2\pi m}{L_x}+\frac{\alpha}{L_x},
\end{align}
with $m = 0, 1, \ldots, \frac{L_x}{a_0} - 1$.
While} only $\alpha=0,\pi$ are allowed {for a class D Hamiltonian, here we consider general $\alpha$ as will be needed for the construction below}.

We now consider the Floquet states in the presence of a time vortex threaded through the holes of the cylinder [as opposed to a point-like time vortex within the bulk of the system as described in Eq.~\eqref{eq:theta}]. 
A time vortex can be threaded through the holes of the cylinder by setting $\theta(\bar{\bm{r}}) = 2\pi \bar{x}/L_x$ in Eq.~(2).
We will show that threading such a time vortex through a hole of the cylinder toggles the presence of the $\pi$ mode in that hole on and off. 
In turn, we will argue that this behavior implies that a point-like time vortex within the bulk of the 2D system must carry a $\pi$ mode.

In Sec.~\ref{app: floquet eigenstates with tv} we show that the Floquet eigenstate in the presence of the time vortex can be written as~\footnote{In the presence of the time vortex, the Hamiltonian is not translationally invariant, but has a combined space-time translation symmetry. The label $k$ of $\chi_{nk}$ is the eigenvalue of the Floquet eigenstate under this combined space-time translation; see Sec.~\ref{app: floquet eigenstates with tv}.} 
\begin{align}
\chi_{nk}(\bm{r},t)=\psi_{nk}(\bm{r},t-\frac{x}{L_x}T;\alpha)+O\left(\frac{\ell}{L_x}\right);
\label{eq: floquet eigenstates with tv on cylinder}
\end{align}
for small $\ell/L_x$, the Floquet states in the presence of the time vortex follow those of the defect-free system, with the local time lag of the Hamiltonian $H_{\rm v}$ imprinted directly onto $\chi_{nk}(\bm{r},t)$. Here, $k$ satisfies Eq.~\eqref{eq:k_quant}. 


Next,
we fix the value of $\alpha$ by imposing the physically desired 
periodic boundary conditions on $\chi_{nk}$:
\begin{align}
\chi_{nk}(\bm{r},t) & =\chi_{nk}(\bm{r}+L_x\hat{\bm{x}},t)\nonumber\\
 & \approx \psi_{nk}(\bm{r}+L_x\hat{\bm{x}},t-\frac{(x + L_x)}{L_x}T;\alpha)\nonumber\\
 & \approx e^{i\alpha}e^{i\varepsilon_{nk}{(\alpha)}T}\chi_{nk}(\bm{r},t),
 \label{eq: chi}
\end{align}
which is satisfied by setting $\alpha$ to $\alpha^\star_{n,m}$, defined as the solution of the equation $\alpha^\star_{n,m}=-\varepsilon_{nk}
(\alpha^\star_{n,m})T$ with $k=(2\pi m + \alpha^\star_{n,m})/L_x$.
This relation must be solved separately for each $n$ and $m$. 
Therefore, comparing with Eq.~(\ref{eq:k_quant}),   
we find that to leading order in $\tfrac{\ell}{L_x}$, the allowed values of $k$ shift as
\begin{align}
k=\frac{2\pi m}{L_x}-\frac{\varepsilon_{nk}(\alpha^\star_{n,m})T}{L_x},
\label{eq:k_time_vortex}
\end{align}
whereas the momentum in the model without the time vortex with periodic
boundary conditions is discretized as $k=\tfrac{2\pi m}{L_x}$. 
In particular, for states near quasi-energy $\pi/T$, 
the insertion of the time vortex effectively changes the boundary conditions 
from periodic to anti-periodic. 
{In contrast, the boundary conditions for states near quasi-energy $\varepsilon_{nk} = 0$ do not change.}

Assuming that there is a chiral mode through quasi-energy $\pi/T$, we can now show that for a finite system with circumference $L_x$, introducing a time vortex through the cylinder toggles between having a Majorana $\pi$ mode at each edge and not having one. 
Linearizing the quasi-energy spectrum near $\varepsilon=\pi/T$ gives
$\varepsilon_{nk}\approx\frac{\pi}{T}+v(k-k_{0})$, where particle-hole symmetry maps $\varepsilon_{k}$ to $-\varepsilon_{-k}$, and hence constrains $k_{0}$ to be either $0$ or $\pi/a_0$. Then, using Eq.~\eqref{eq:k_quant} with $\alpha=0$ and Eq.~\eqref{eq:k_time_vortex} to describe systems with $n_v=0$ or 1 time vortices, respectively, we find that the quasi-energy spectrum for a finite system, to order $\ell/L_x$, is given by
\begin{align}
    \varepsilon_{nk}=v\left(\frac{\pi\left(2m - n_{v}\right)}{L_{x}}-k_{0}\right)+\frac{\pi}{T}.
    \label{eq:finite system spectrum}
\end{align}
 Using Eq.~\eqref{eq:finite system spectrum}, we can verify that if the system has a Majorana $\pi$ mode at the edge (whose quasi-energy is exactly $\pi/T$) in the absence of a time vortex ($n_v=0$), then the system with $n_v=1$ does not have $\pi$ mode, and vice versa \footnote{In particular, if $k_0=0$, the system with $n_v=0$ has a $\pi$ mode at the edges and the $n_v=1$ system does not. If $k_0=\pi/a_0$, the $n_v=0$ system has a $\pi$ mode if $L_x/a_0$ (the number of sites around the cylinder) is even, and does not have one if $L_x/a_0$ is odd. In the $n_v=1$ case, the situation is reversed.}.


This argument implies that a point-like time vortex within the bulk must carry a $\pi$ mode. To see this, consider a cylinder with a single time vortex piercing through its surface. Then, there must also be a time vortex through \emph{one} of the holes of the cylinder but not the other. 
According to Eq.~\eqref{eq:finite system spectrum}, adding a time vortex through a hole toggles between having a $\pi$ mode or not at the corresponding edge. Therefore, inserting a time vortex through
one the holes of the cylinder and extracting it through the bulk 
changes the parity of the number of 
$\pi$ modes on the edges from even to odd. 
Since the total number of Majorana $\pi$ modes in the entire system must be even, the other Marjorana $\pi$ mode has to be located in the bulk; this $\pi$ mode is localized around the core of the bulk time vortex, which is the only point where the bulk gap around quasi-energy $\pi/T$ can close.

\subsection{Periodically driven Kitaev honeycomb model}
\label{sec: Kitaev}

As an application of these ideas, we use a periodically driven Kitaev spin model on the honeycomb lattice \cite{Fulga2019}, which maps to a problem of free fermions in class D (Majorana fermions) coupled to a static background $\mathbb{Z}_2$ gauge field \cite{Kitaev2006}.
Introducing a time vortex enriches the types of topological excitations that are accessible in this system. In addition, the spin model and the associated time vortices can be realized on near-term digital quantum computers.

\begin{figure}
\begin{centering}
\includegraphics[trim={0cm 2.8cm 1.9cm 2.7cm},clip,width=(\textwidth-\columnsep)/2]{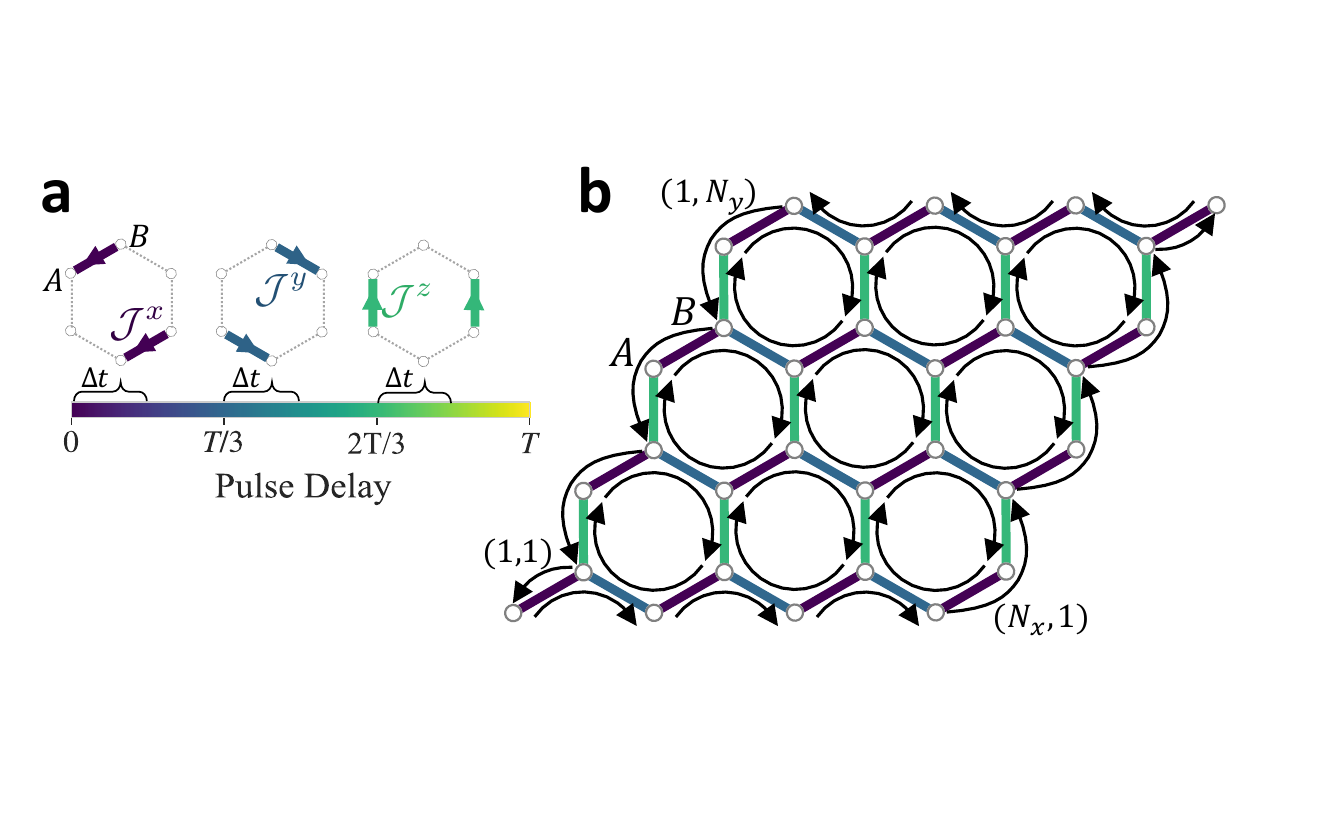}
\par\end{centering}
\caption{\textbf{The dynamics of the driven Kitaev honeycomb model.} \textbf{(a)} The Kitaev honeycomb model in terms of the fermionic degrees of freedom includes nearest neighbor hopping terms with a strength dependent on the orientation of the bonds $\mathcal{J}^{a=x,y,z}$. In the periodically driven version of the model, these three terms are applied in different time intervals $[lT+t_a,lT+t_a+\Delta t]$. The arrows indicate a gauge choice for $u_{\mathbf{I},\mathbf{J}}$ that corresponds to $W_{p}=1$ (no $\mathbb{Z}_2$ flux); within this gauge choice, hopping of a $c$ fermion from sublattice $B$ to $A$ gives a phase $i$. \textbf{(b)} In the limit of resonant driving, $\mathcal{J}^a_0 \Delta t=\pi/4$, the action of the full driving period on the Majoranas is a permutation which is illustrated by black arrows. This yields flat bulk bands at quasienergy $\varepsilon=\pm\pi/(2T)$ and chiral edge states in a finite system.}
\label{fig: floquet honeycomb model without time vortex}
\end{figure}

The model consists of spin-$\frac{1}{2}$ degrees of freedom on the vertices of a honeycomb lattice with time dependent 
anisotropic exchange interactions. The Hamiltonian is
\begin{align}
\mathcal{H}(t)=&-\sum_{a\in\{x,y,z\}}\sum_{\langle \mathbf{I},\mathbf{J}\rangle\in a\text{-bonds}}\mathcal{J}^{a}_{\mathbf{I},\mathbf{J}}(t)\sigma_{\mathbf{I}}^{a}\sigma_{\mathbf{J}}^{a}.
\end{align}
Here, $\mathbf{I} = (i,j,s)$ denotes the sites of the honeycomb lattice, where $i \in \{1,\dots,N_x\}$ and $j \in \{1,\dots,N_y\}$ label a unit cell, 
and $s=A,B$ labels the two sublattices. Each site hosts a spin $\bm{\sigma}_{\mathbf{I}}$. The bonds of the lattice are partitioned into three sets $a = \{x, y, z\}$ according to the three types of bond orientations in the honeycomb lattice. 
We take $\mathcal{J}^{a=x,y,z}_{\mathbf{I},\mathbf{J}}$ to be periodically time dependent with a period $T$ [see Fig.~\ref{fig: floquet honeycomb model without time vortex}a]. 

This model can be elegantly solved with arbitrary coupling strengths by fermionization as is done in Ref.~\cite{Kitaev2006}. We introduce a set of four Majorana operators $\{b_{\mathbf{I}}^{x},b_{\mathbf{I}}^{y},b_{\mathbf{I}}^{z},c_{\mathbf{I}}\}$ for each site $\mathbf{I}$, and a constraint $b_{\mathbf{I}}^{x}b_{\mathbf{I}}^{y}b_{\mathbf{I}}^{z}c_{\mathbf{I}}=1$ which results in a two level system for each site. 
We map the spins to fermions via $\sigma_{\mathbf{I}}^{a} = ib_{\mathbf{I}}^{a}c_{\mathbf{I}}$ and identify a conserved $\mathbb{Z}_2$ gauge field defined on the bonds of the lattice,
\begin{align}
\left\{u_{\mathbf{I},\mathbf{J}}=ib_{\mathbf{I}}^{a}b_{\mathbf{J}}^{a}|\langle \mathbf{I},\mathbf{J}\rangle\in a\text{-bonds},a\in\{x,y,z\}\right\}.
\end{align}
This leads to the Hamiltonian
\begin{align}
\mathcal{H}(t)=-\sum_{a\in\{x,y,z\}}\sum_{\langle \mathbf{I},\mathbf{J}\rangle\in a\text{-bonds}}\mathcal{J}^{a}_{\mathbf{I},\mathbf{J}}(t)u_{\mathbf{I},\mathbf{J}}ic_{\mathbf{I}}c_{\mathbf{J}}.
\label{eq: KSL Hamiltonian in terms of fermions}
\end{align}
The 
variables $u_{\mathbf{I},\mathbf{J}}$ 
are not gauge-invariant: applying the gauge operator $b_{\mathbf{I}}^{x}b_{\mathbf{I}}^{y}b_{\mathbf{I}}^{z}c_{\mathbf{I}}$ changes 
the sign of $u_{\mathbf{I},\mathbf{J}}$ on the three bonds touching vertex $\mathbf{I}$. However, the $\mathbb{Z}_2$ flux $W_{p}$ given by the product of the gauge fields around a hexagonal plaquette $p$ is gauge-invariant. 
Note also that the free fermion Hamiltonian \eqref{eq: KSL Hamiltonian in terms of fermions} is purely imaginary, and is hence particle-hole symmetric with $\mathcal{P}=\mathbb{I}$.


Using the fact that the 
operators $\{u_{\mathbf{I},\mathbf{J}}\}$ are conserved by the Hamiltonian, we fix a gauge and replace them with their eigenvalues, $u_{\mathbf{I},\mathbf{J}}=\pm1$. With this substitution, the Hamiltonian \eqref{eq: KSL Hamiltonian in terms of fermions} describes free fermions, $\{c_{\mathbf{I}}\}$.

We start from a translation invariant model given by $\mathcal{J}^{a}_{\mathbf{I},\mathbf{J}}(t)=\tilde{\mathcal{J}}^{a}(t)$, where we choose each $\tilde{\mathcal{J}}^a(t)$ to be a train of rectangular pulses of width $\Delta t$ and amplitude $\mathcal{J}^a_0$, starting at time $lT+t_a$ and ending at $lT+t_a+\Delta t$, for $l\in\mathbb{Z}$. 
We take $t_x=0$, $t_y=\frac{1}{3}T$, $t_z=\frac{2}{3}T$.
The Kitaev honeycomb model is illustrated in Fig.~\ref{fig: floquet honeycomb model without time vortex}.


In order to insert a time vortex at $\bm{r}_0$, we follow the prescription in Sec.~\ref{sec: time vortex - general} by adding a delay as a function of the position of the midpoint of each bond 
$\bar{\bm{r}}_{\mathbf{I},\mathbf{J}}=\frac{1}{2}\left(\bm{r}(\mathbf{I})+\bm{r}(\mathbf{J})\right)$ such that

\begin{align}
\mathcal{J}^{a}_{\mathbf{I},\mathbf{J}}(t)=\tilde{\mathcal{J}}^{a}\left(t-T\frac{\theta\left(\bar{\bm{r}}_{\mathbf{I},\mathbf{J}}\right)}{2\pi}\right).
\end{align}

In a time-reversal symmetry broken version of the static Kitaev model, a phase with a nonzero Chern number of the $c_{\mathbf{I}}$ fermions can occur in which $\mathbb{Z}_2$ fluxes (where $W_{p} = -1$) 
bind Majorana zero modes. In contrast, Ref.~\cite{Fulga2019} showed that in the anomalous Floquet phase, realized in the vicinity of $\mathcal{J}^a_0\Delta t=\pi/4$, such $\mathbb{Z}_2$ fluxes carry Majorana modes at \textit{both} $0$ and $\pi/T$ quasienergies. Sec.~\ref{app: phase diagram} presents the phase diagram of the Floquet Kitaev honeycomb model.

The limit of short pulses with no overlaps, $\Delta t<T/3$, at ``resonance,'' $\mathcal{J}^a_0\Delta t=\pi/4$, turns out to be very simple to analyze. Each pulse serves to swap the Majorana operators connected by the bond on which it acts. After the action on the bond $\langle \mathbf{I},\mathbf{J}\rangle$ with $\mathbf{I}$ in the $A$ sublattice and $\mathbf{J}$ in the $B$ sublattice, $c_\mathbf{I}\rightarrow c_\mathbf{J}$ and $c_\mathbf{J}\rightarrow -c_\mathbf{I}$. In the spatially uniform model, the order of the pulses dictates that a full period of time evolution effectively swaps a pair of Majoranas located on opposite 
corners of each plaquette in the bulk, $c_{(i,j,A)}\rightarrow-c_{(i+1,j-1,B)}$ and $c_{(i+1,j-1,B)}\rightarrow c_{(i,j,A)}$. The result is that Floquet eigenmodes are 
split across pairs of sites as $\left[c_{(i,j,A)}\pm ic_{(i+1,j-1,B)}\right]/\sqrt{2}$, with quasienergies $\pm\tfrac{\pi}{2T}$.
Creating a $\mathbb{Z}_2$ flux on a plaquette adds a minus to the exchange of the two fermions, altering the  Floquet eigenmodes to real 
combinations of the form $\left[c_{(i,j,A)}\pm c_{(i+1,j-1,B)}\right]/\sqrt{2}$, with quasienergies $\varepsilon=0,\tfrac{\pi}{T}$. In terms of the spin degrees of freedom, at the resonant limit each pulse simply implements a Clifford gate on neighboring spins $U=\exp(i\frac{\pi}{4}\sigma_{\mathbf{I}}^{a}\sigma_{\mathbf{J}}^{a})$, where we note that an operator of the form $\exp({i\frac{\pi}{4}P})$ is a Clifford unitary for any (multi-qubit) Pauli operator $P$.


\subsubsection{Time vortex}
\label{sec: resonant limit}

\begin{figure}
\begin{centering}
\includegraphics[trim={1.6cm 1.6cm 1.8cm 2.6cm},clip,width=(\textwidth-\columnsep)/2]{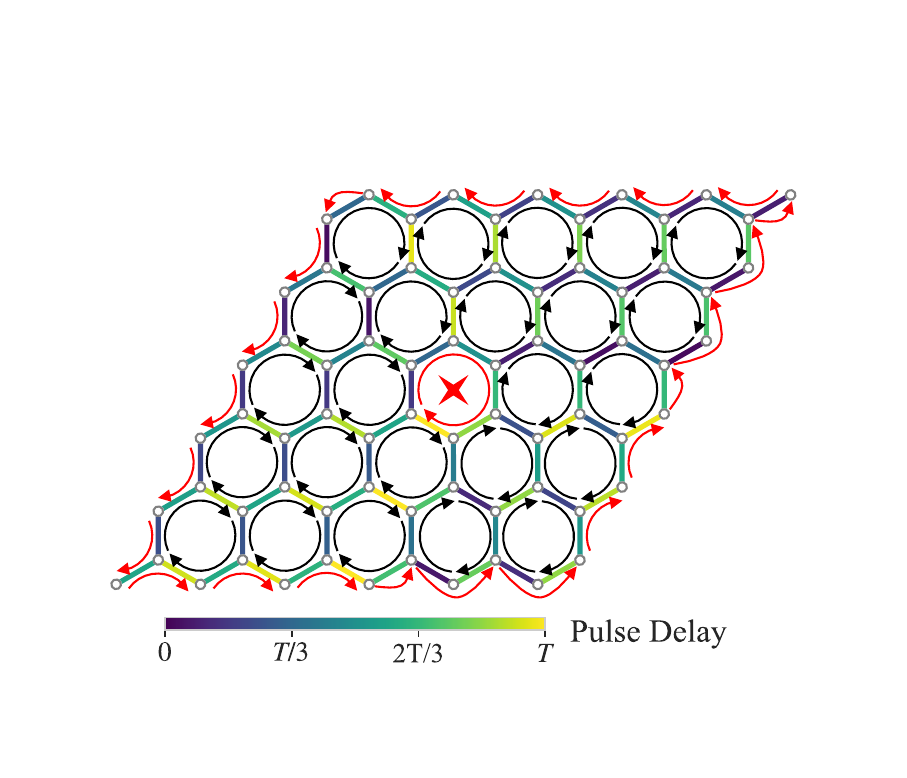}
\par\end{centering}
\caption{\textbf{The driven Kitaev honeycomb model with a time vortex at the middle plaquette in the resonant instantaneous pulse limit.} The bonds are colored according to the time at which they act in units of the period $T$. The action of the full driving period on the Majoranas is a permutation which is illustrated by arrows. A single Majorana mode (red) at the time vortex core returns to itself with a minus sign; this is a bound $\pi$ mode. Another $\pi$ mode appears in the spectrum of the chiral edge state of the system as can be seen by the fact that the cycle of the Majorana modes on the edge (red) must also be of odd length.}
\label{fig: pulse order}
\end{figure}

Now we consider inserting a time vortex at the center of a plaquette 
in a system with no $\mathbb{Z}_2$ fluxes. 
As described above, the time vortex adds delays to the time dependence of the couplings of each bond relative to the uniform system. 
The simplicity of the analysis above for the uniform case with short pulses was enabled by the fact that the exchange pulses on neighboring bonds never acted simultaneously (allowing disconnected pairs of sites to evolve within each pulse).
This simplicity can be retained for the analysis of a situation with a time vortex by taking the pulses to be sufficiently short, with $\Delta t<T/6$, such that the pulses acting on the bonds of the plaquette 
hosting the time vortex core are sufficiently separated in time that no overlaps are created. 
We find it furthermore convenient to take $\Delta t\lesssim T/N_x$, $\Delta t\lesssim T/N_y$, so that at the initial time $t=0$ none of the pulses anywhere on the lattice are active. In this ``instantaneous pulse'' limit, the effect of the time vortex is only a rearrangement of the order of the pulses acting on the bonds of the lattice. The order of the pulses acting on the bonds and the action of the full period evolution operator on the Majoranas is illustrated in Fig.~\ref{fig: pulse order}.
Below we will demonstrate with numerical simulations that the qualitative results we obtain from this simple limit persist under more generic driving conditions, including the case $\Delta t > T/3$ where neighboring pulses overlap.

It is easy to see that in a region far away from the vortex core, the order of pulses acting on the bonds will locally be the same as in the absence of the vortex. However, depending on the orientation of the region relative the vortex core, the order may change by a cyclical permutation to become $y\rightarrow z\rightarrow x$ or $z\rightarrow x\rightarrow y$. In any case, plaquettes in such regions will host localized Floquet modes with quasienergies $\pm\frac{\pi}{2T}$. 
Thus the bulk spectrum 
remains undisturbed by the insertion of the time vortex.

The effect of the time vortex at the vortex core is also easy to understand. The order of the pulses on the bonds around the plaquette hosting the time vortex core becomes a cycle going around the plaquette in the clockwise direction, starting from the $z$ pulse on the left bond of the plaquette (Note that all six pulses around the core act within one driving period $T$). 
Therefore, the Majorana at the lower left corner of the plaquette $c_{(i,j,B)}$ will hop entirely around the plaquette in a single period. To determine the quasi-energy of the corresponding localized mode, we note that each pulse implements a transformation $c_A\rightarrow c_B$, $c_B\rightarrow -c_A$. Following the arrows in Fig.~\ref{fig: floquet honeycomb model without time vortex}a, three factors of $(-1)$ are incurred over the entire period.
Consequently, the overall phase is $\pi$, and we find that this eigenstate is a {Majorana $\pi$ mode} bound to the time vortex. 

\subsubsection{$\pi$ modes in a finite size simulation}
\label{sec: finite size numerics}

After confirming the existence of the $\pi$ mode bound to the time vortex in the resonant, instantaneous pulse limit, we now consider the generic situation where the pulses do not satisfy the resonance condition and 
pulses on neighboring bonds overlap in time. We demonstrate that the $\pi$ mode survives and remains localized around the vortex core. The phase diagram of the Floquet Kitaev honeycomb model highlighting the region in parameter space in which a time vortex binds a $\pi$ mode is shown in Sec.~\ref{app: phase diagram}.

{To treat the generic situation,} we numerically perform time-evolution of the Majorana operators $c_\mathbf{I}$ in the Heisenberg picture. 
We then calculate the spatially resolved time-averaged density of states \cite{rudner2020floquet, Uhrig_2019} 
\begin{align}
\bar{\rho}_\mathbf{I}(\omega)=-\frac{1}{\pi}\operatorname{Im}{\widetilde{G}^{R}_{\mathbf{I},\mathbf{I}}(\omega)}.
\end{align}
Here $\widetilde{G}^{R}_{\mathbf{I},\mathbf{J}}(\omega)$ is the Fourier transform of the time-averaged single-particle retarded Green’s function given by
\begin{align}
\bar{G}^{R}_{\mathbf{I},\mathbf{J}}(t)=\frac{1}{T}\int_{0}^{T}d\tau G^{R}_{\mathbf{I},\mathbf{J}}(t+\tau,\tau),
\end{align}
where the Green's function itself is
\begin{align}
G^{R}_{\mathbf{I},\mathbf{J}}(t+\tau,\tau)=-\frac{i}{4}\theta(t)\left<\{c_\mathbf{I}(t+\tau),c_\mathbf{J}(\tau)\}\right>.
\end{align}
Here, since the $c_{\mathbf{I}}$ fermions are free, the expectation value is independent of the state it is taken in, as long as there are no $\mathbb{Z}_2$ fluxes in the system.

\begin{figure}
\begin{centering}
\includegraphics[trim={0cm 5.2cm 1.8cm 3cm},clip,width=(\textwidth-\columnsep)/2]{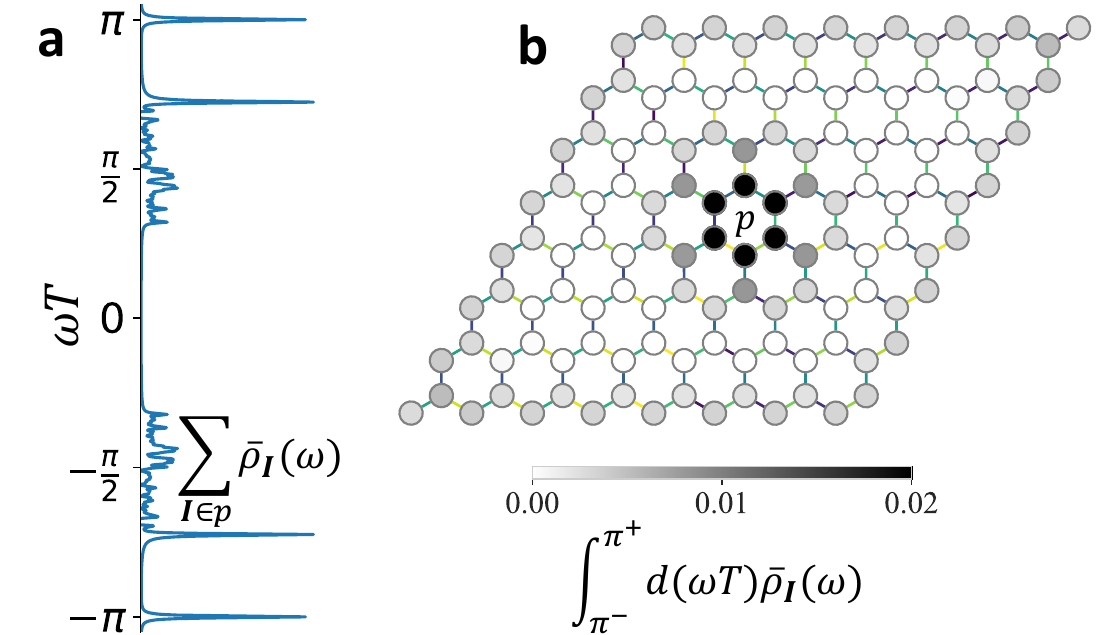}
\par\end{centering}
\caption{\textbf{The driven Kitaev honeycomb model with a time vortex away from resonance.} The time-averaged density of states of the driven Kitaev honeycomb model with a time vortex at the middle plaquette, with overlapping and non-resonant pulses: 
$\Delta t=0.5T$, $\mathcal{J}^a_0\Delta t=0.9\pi/4$. Panel \textbf{(a)} shows this density of states summed over the sites nearest the vortex core. In panel \textbf{(b)} the density of states near energy $\pi/T$ is plotted on the lattice.}
\label{fig: spectrum}
\end{figure}

Figure~\ref{fig: spectrum}a shows the time-averaged density of states summed over the sites of the central plaquette (which hosts the time vortex), in a system of size $16\times16$ unit cells. The parameters of the simulation are given in the figure caption. 
In this generic regime, the flat bulk bands at quasi-energy $\pm\pi/(2T)$ 
{discussed above for the} resonant limit acquire a finite width, as expected. 
{Importantly,} the gaps around quasi-energy $0$ and $\pi/T$ remain finite. 
Consequently, the $\pi$ Majorana mode at the vortex core remains robust, as can be seen by the sharp peak at $\omega=\pi/T$.
{We also find an additional} (unprotected) bound state at the vortex core, near the top of the upper band and the bottom of the lower band. 

Figure \ref{fig: spectrum}b shows the spatially resolved time-averaged density of states, integrated within the range $\omega\in [0.99\pi/T,1.01\pi/T]$, in a system of size $8\times8$ unit cells. The spectral function has support near the center, where the time vortex is located (the plaquette indicated by $p$ in the figure), and at the edge of the system. The $\pi$ mode at the vortex core hybridizes with the $\pi$ mode at the edge, but this hybridization is expected to decrease exponentially with the system size.


\subsubsection{Topological invariant}
\label{sec: invariants}


Topological defects in periodically driven non-interacting systems of different symmetry classes have been classified in Ref.~\cite{Yao2017}. In particular, it was found that a zero-dimensional (point-like) defect in a two-dimensional periodically driven system is characterized by a $\mathbb{Z}_2\times \mathbb{Z}_2$ invariant. The two $\mathbb{Z}_2$ invariants, $\nu_0$ and $\nu_\pi$, are equal to the number of Majorana $0$ and $\pi$ modes bound to the center of the defect, respectively. 
Our goal is to compute $\nu_0$ and $\nu_\pi$ for the time vortex; in particular, we show that in the anomalous phase of the driven Kitaev honeycomb model of Sec.~\ref{sec: Kitaev}, $\nu_0=0$ and $\nu_\pi=1$. 

We consider the Hamiltonian acting on a region far from the time vortex core, at an angle $\theta$ around the core [as defined in Eq.~(\ref{eq:theta})]. The Hamiltonian is a slowly varying function of the position $\bar{\bm{r}}$, so by treating $\bar{\vec{r}}$ as a constant in a given region we can define the Bloch Hamiltonian $H_0(\bm{k},t;\theta)$ \cite{Teo2010}, obtained by taking $\theta$ in the right hand side of Eq. \eqref{eq:H_theta} to be constant and Fourier transforming with respect to $\bm{r}-\bm{r}'$. As defined, $H_0(\bm{k},t;\theta)$ is periodic both in $t$ and in $\theta$. 
A construction of the $\mathbb{Z}_2\times \mathbb{Z}_2$ invariant for $H_0(\bm{k},t;\theta)$ was given in Ref.~\cite{Yao2017}; here, we give an alternative formulation of the invariant, which is more directly related to the number of Majorana modes at the defect. 

\begin{figure}
\begin{centering}
\includegraphics[trim={0cm 0cm 0cm 0cm},clip,width=(\textwidth-\columnsep)/2]{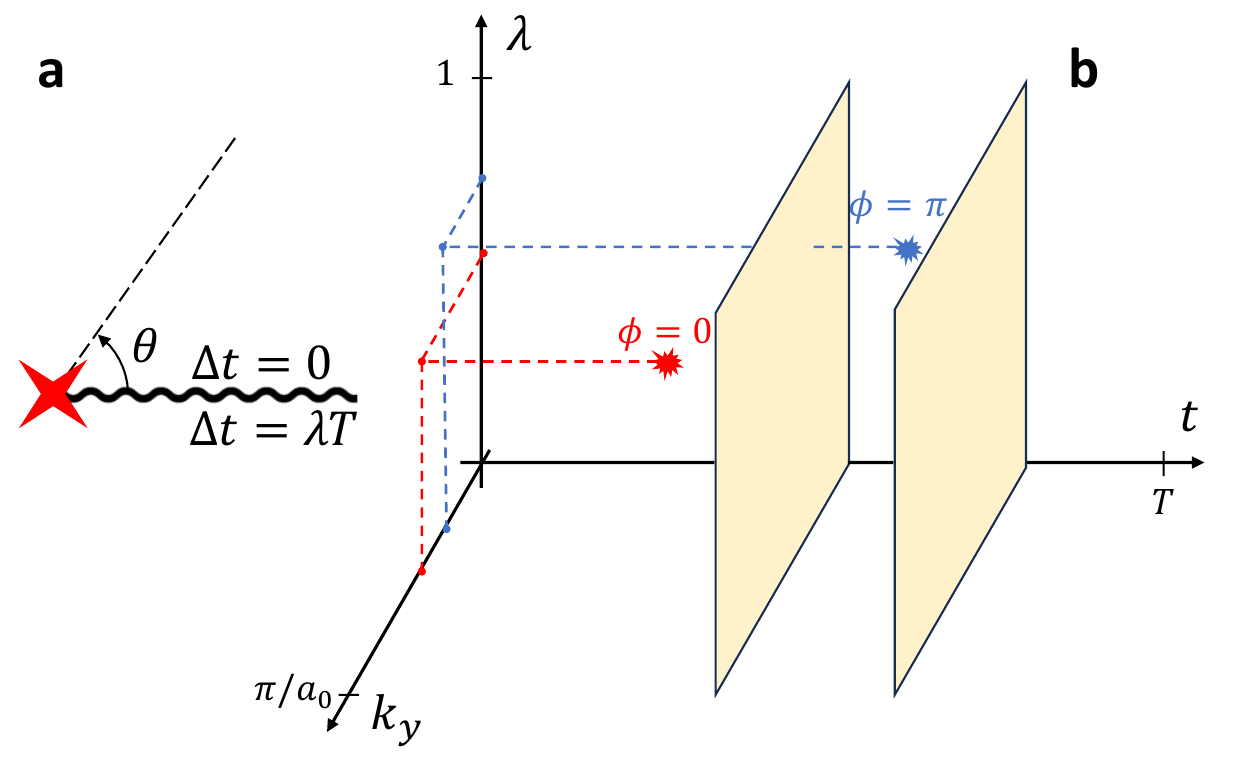}
\par\end{centering}
\caption{\textbf{Computing the topological invariants of the time vortex defect.} \textbf{(a)} A family of Hamiltonians parameterized by $\lambda\in[0,1]$ interpolates between a translation invariant model (for $\lambda=0$) and a model with a time vortex ($\lambda=1$) by gradually increasing the time delay of the driving Hamitonian according to Eq. \eqref{eq:H_lambda}. This provides a mapping to a 1D problem at the branch cut discontinuity at $\theta=0$. \textbf{(b)} 
To determine the topological properties of the time vortex, we examine the spectrum of $H_{\rm{eff}}(\bm{k},t;2\pi\lambda)$ [Eq. \eqref{eq:H_eff}] for fixed $k_x=0$ or $\tfrac{\pi}{a_0}$, as a function of $k_y$, $t$, and $\lambda$. 
The individual topological invariants at quasienergy $\varepsilon=0$ and $\varepsilon=\pi/T$ may change at topological singularities of the phase bands at corresponding phases which are illustrated by red and blue symbols. Computing the topological invariant of an effective non-driven model on time-like surfaces in the $k_y,\lambda,t$ space (yellow) before and after each singularity allows us to resolve the two separate invariants of the full model.}
\label{fig: schematic topological singularities}
\end{figure}

To this end, we construct a family of time-dependent Hamiltonians parameterized by $\lambda\in[0,1]$ which smoothly interpolates between the vortex-free model, obtained for $\lambda=0$ (with trivial topological invariants) and the model with the vortex for $\lambda=1$:
\begin{align}
H(\bm{k},t; \theta,\lambda) = H_0(\bm{k},t; \theta\lambda).
\label{eq:H_lambda}
\end{align}
This family of Hamiltonians is not periodic in $\theta$ for $\lambda\notin\{0,1\}$. 
The non-periodicity has a simple consequence in real space: the position-dependent Hamiltonian around the time vortex changes abruptly along the ray $\theta=0$ unless $\lambda=0$ or 1 [Fig.~\ref{fig: schematic topological singularities}(a)]. One-dimensional states that are bound to this ray may close the Floquet gaps around quasi-energy $0$ and $\pi/T$ at some values of $\lambda$.
Such a gap closing and reopening, corresponding to a topological phase transition along the $\theta=0$ ray, is necessary in order to obtain a Majorana mode at the core of the defect (the edge of the $\theta=0$ ray) for $\lambda=1$. 

Thus, we have reduced the problem to determining the topological properties of a one-dimensional 
system along the $\theta=0$ ray in real space. Far away from the vortex core,  the $\theta=0$ ray can be viewed as a domain wall between a $y>0$ region   
where the phase of the driving Hamiltonian is $0$ 
and a $y<0$ region where the phase is $2\pi \lambda$  (see Fig.~\ref{fig: schematic topological singularities}a). 
The system with such a domain wall is translationally invariant along the $\theta=0$ ray, and hence $k_x$ is conserved.


As is often the case for systems in class D, to probe for topological gap closings and re-openings in the $\theta=0$ ray as a function of $\lambda$, it suffices to consider the two particle-hole symmetric momenta, $k_x=0$ and $\pi/a_0$ \cite{Kitaev2001}. 
This can be done by studying $H_0(k_x\in\{0,\tfrac{\pi}{a_0}\},k_y,t;2\pi\lambda)$, which describes the region below the domain wall ($y<0$). 
For fixed values of $k_x$, this Hamiltonian describes a family of one-dimensional (along $k_y$) Floquet Hamiltonians parametrized by $\lambda$. 
Each such family of 1D Hamiltonians is itself characterized by a pair of $\mathbb{Z}_2$ invariants~\cite{Yao2017}, one for each of the particle-hole symmetric quasi-energies, $\varepsilon=0$ and $\pi/T$. 

Physically, $H_0(k_x\in\{0,\tfrac{\pi}{a_0}\},k_y,t;2\pi\lambda)$ defines a Floquet version of an ``adiabatic fermion parity pump''~\cite{Teo2010,Keselman2013}. 
We denote the resulting $\mathbb{Z}_2$ invariants as $\nu(\varepsilon, k_x)$, corresponding to the fermion parity pumped by the system labeled by $k_x\in\{0,\tfrac{\pi}{a_0}\}$ at quasi-energy $\varepsilon\in\{0,\tfrac{\pi}{T}\}$ as the parameter $\lambda$ changes adiabatically from $0$ to $1$. We identify the topological invariants of the time vortex as $\nu_0 = \nu(0, 0) + \nu(0, \tfrac{\pi}{a_0}) \pmod{2}$ and $\nu_\pi = \nu(\tfrac{\pi}{T}, 0)  + \nu(\tfrac{\pi}{T}, \tfrac{\pi}{a_0}) \pmod{2}$ \footnote{If $\nu(\varepsilon, k_x)=1$, then the gap at the end of the system has to close at quasi-energy $\varepsilon$ for some value of $\lambda$. This, in turn, implies a topological gap closing along the ray at the same $k_x$ at this value of $\lambda$ (i.e., a level crossing of two states related by particle-hole symmetry).
Viewed as an adiabatic process, such a level crossing changes the fermion parity corresponding to the Dirac fermion formed by the two Majorana states that cross, hence it is referred to as a fermion parity pump.
}. 

The invariants $\nu(\varepsilon,k_x)$ can be determined as follows. We examine the evolution operator from time $0$ to $t$,
\begin{equation}    U(\bm{k,}t;2\pi\lambda)=\mathcal{T}\exp\left[-i\int_{0}^{t}d\tau H_0(\bm{k},\tau;2\pi\lambda)\right],
\end{equation}

in the three-dimensional space of $k_y\in [0,\tfrac{\pi}{a_0}]$,  $\lambda\in [0,1)$, and $t\in[0,T)$, for fixed $k_x=0$ or $\tfrac{\pi}{a_0}$ (Fig.~\ref{fig: schematic topological singularities}b). 
We denote the phases of the eigenvalues of $U$ (known as the phase bands~\cite{Nathan2015}) by $\phi_n(\bm{k},t;2\pi\lambda)$, where $n$ is the band index. 
Along the plane $t=0$, we have $U=\mathbb{I}$. 
At $t=T$, $U(\bm{k},T;2\pi\lambda)$ is equal to the Floquet operator, whose spectrum is assumed to be gapped around $\varepsilon=0$ and $\pi/T$. 

For every plane of fixed time $t>0$ for which the spectrum of $U(k_x\in \{0,\tfrac{\pi}{a_0}\},k_y,t; 2\pi\lambda)$ is gapped (i.e., the phase bands do not cross $0$ and $\pi$), 
a $\mathbb{Z}_2$ invariant can be defined as follows. We consider the effective Hamiltonian 
\begin{align}
    H_{{\rm eff}}(\bm{k},t;2\pi\lambda)=\frac{i}{t}\log[U(\bm{k,}t;2\pi\lambda)],
    \label{eq:H_eff}
\end{align}
for fixed $k_x\in\{0,\tfrac{\pi}{a_0}\}$, as a function of $k_y$ and $\lambda$~\footnote{Since $H_{\rm{eff}}(\bm{k},t;2\pi\lambda)$ is assumed to be gapped at $\varepsilon=\tfrac{\pi}{T}$ for this value of $t$, we can use the usual convention for the branch cuts of $\log(z)$, with a discontinuity along $\rm{Re(z)<0}$, $\rm{Im}(z)=0$}. This can be viewed as a time-independent 1D Hamiltonian in class D that depends periodically on a parameter $\lambda$; the $\mathbb{Z}_2$ invariant is equal to the fermion parity pumping over a single cycle of changing $\lambda$ adiabatically~\cite{Teo2010,Keselman2013}. For an explicit prescription for the computation of this invariant, see Sec.~\ref{app: Topological invariants resolved on time-like surfaces}. 

The sum of the invariants at quasi-energies $0$ and $\pi/T$, $\nu(0,k_x) + \nu(\pi/T,k_x) \pmod{2}$, is simply equal to the fermion parity pumping invariant of $H_{\rm{eff}}(\bm{k},T;2\pi\lambda)$ as a function of $k_y$ and $\lambda$ for fixed $k_x\in \{
0,\tfrac{\pi}{a_0}\}$. Disentangling $\nu(0,k_x)$ and $\nu(\tfrac{\pi}{T},k_x)$ is more subtle, and depends on the evolution operator at intermediate times, $0<t<T$. To this end, we note that $\nu(\pi/T,k_x)$ can only change via a topological gap closing of $H_{\rm{eff}}(\bm{k},t,2\pi\lambda)$ at $\varepsilon=\pi/T$ for some intermediate $t$. We therefore track the topological singularities where the gap of 
$H_{\rm{eff}}(\bm{k},t;2\pi\lambda)$ closes at quasi-energy $\pi/T$~\cite{Nathan2015}, and compute the $\mathbb{Z}_2$ fermion parity pumping invariants of $H_{\rm{eff}}(\bm{k},t;2\pi\lambda)$ on constant $t$ planes 
before and after the singularities, illustrated in Fig.~\ref{fig: schematic topological singularities}b. 
For each such pair of planes, we define an index $\zeta_i$ equal to the difference (mod 2) of the $\mathbb{Z}_2$ invariants of the two planes, that captures the change in the invariant across the singularity.
The sum (mod 2) of $\zeta_i$'s across all $\varepsilon=\tfrac{\pi}{T}$ topological singularities for $0<t<T$ gives $\nu(\tfrac{\pi}{T},k_x)$. 

In practice, the topological singularities may occur on higher-dimensional regions (lines, surfaces, or volumes) in the $(k_y,t,\lambda)$ space, rather than at isolated points (as assumed, for simplicity, in Fig.~\ref{fig: schematic topological singularities}b). 
In that case, there may not be planar surfaces that separate the $\varepsilon=0$ and $\tfrac{\pi}{T}$ singularities. 
To handle this situation, $U(\bm{k},t;2\pi\lambda)$ can in principle be deformed continuously without closing the gap at $t=T$ such that there are planes of constant $t$ that separate all the topological singularities. 
Alternatively, one can find non-planar surfaces that separate the singularities, on which the invariants of $H_{\rm{eff}}$ are computed;  
these surfaces are required to be periodic in $\lambda$ for the invariant to be well-defined. 

Using this procedure, we find the invariants of the time vortex in the anomalous phase of the Kitaev honeycomb model to be $\nu_0=0$, $\nu_\pi=1$, as expected. The details of the computation are provided in Sec.~\ref{app: Topological invariants resolved on time-like surfaces}. The topological invariants of the two types of topological defects in this phase, the $\mathbb{Z}_2$ flux and the time vortex, are summarized in Table \ref{tab: invariants}.

\begin{table}[]
    \caption{The invariants at quasienergy $0$ and $\pi/T$ of topological defects in 2D.}
    \label{tab: invariants}
    \centering
    \begin{tabular}{c|c|c|c|c}
         &Trivial & $\mathbb{Z}_2$ Flux & Time Vortex & Both \\
         \hline
         $0$ & $0$ & $1$ & $0$ & $1$ \\
         $\pi/T$ & $0$ & $1$ & $1$ & $0$
    \end{tabular}
\end{table}

\section{Discussion}

To summarize, in this work, 
we have considered topological defects in periodically driven systems that utilize their intrinsic symmetry under translation in time. 
A spatially uniform shift in the origin of time does not affect the Floquet spectrum, and changes the Floquet states by a unitary transformation. 
In contrast, a space-dependent time delay of the drive can have physical consequences. 
Intuitively, a time vortex, where this time delay winds by a full period, has a simple effect on the Floquet states far away from the center of the defect: when going around the defect, the wavefunction acquires a phase which is proportional to the quasi-energy of the Floquet state times $T$. This phase can be thought of as arising from the additional driving period ``experienced'' by the wavefunction upon encircling the defect. As we have shown, in class D, the change of phase implies that a time vortex binds a Majorana $\pi$ mode. More generally, a time vortex defect with $n_v$-fold winding binds a Majorana $\pi$ mode if $n_v$ is odd.

We note that the appearance of topological $\pi$ modes at time vortices is not unique to symmetry class D. 
Similar phenomena are expected to occur in any symmetry class whose topological classification in $d=2$ is non-trivial (such that topological edge states exist), and that includes
particle-hole symmetry (such that $\pi$ modes can be topologically protected). In such cases, if the gap at quasi-energy $\pi/T$ is non-trivial, a time vortex may bind $\pi$ modes. In addition to class D, these conditions are met also in classes DIII and C of the Altland-Zirnbauer classification of topological insulators and superconductors~\cite{Nathan2015,Yao2017}.


In terms of practical utility, this construction may be useful for topological quantum computation as it allows manipulating logical qubits encoded in zero and $\pi$ modes independently. 
Using both types of Majorana modes 
leads to a more efficient encoding rate while still protecting the logical qubits from local noise, as long as the noise varies slowly in time, 
since the logical operations involve pairs of Majorana operators separated in space or in quasienergy. 


The time vortex is an externally imposed topological point defect and can be toggled and manipulated by the driving Hamiltonian. 
In contrast, the $\mathbb{Z}_2$ flux is an intrinsic excitation which depends on the quantum state rather than on the driving Hamiltonian, perhaps making it more challenging to control. 
In fact, as demonstrated in Sec.~\ref{sec: resonant limit}, the time vortex can be implemented in a very simple spin model requiring only two-body local Clifford gates, prepared in the $\mathbb{Z}_2$ flux-free sector (topologically equivalent to the ground state of the toric code). 
The time vortex can be toggled on and off simply by rearranging the order of the periodic gate sequence. This makes the time vortex particularly suitable for implementation on quantum devices which allow for a high accuracy application of Clifford gates. 





\section{Methods}

\subsection{Approximate Floquet states with a time vortex}
\label{app: floquet eigenstates with tv}
In this section we explicitly verify that the ansatz for the Floquet states on the cylinder in the presence of the time vortex in Eq.~\eqref{eq: floquet eigenstates with tv on cylinder} satisfies the Schr\"{o}dinger equation
to leading order in $\frac{\ell}{L_x}$. 
By substitution we find
\begin{align}
    &\left[i\partial_{t}  -H_{{\rm v}}(t)\right]  \chi_{nk}(t) \nonumber \\
    &= \left[i\partial_{t}-H_{{\rm 0}}\left(t-\frac{(x+x')T}{2L_x}\right)\right]\psi_{nk}(t-\frac{x'}{L_{x}}T;\alpha)\nonumber\\
    &+O\left(\frac{\ell}{L_x}\right),
    \label{eq:time dependent schrodinger}
\end{align}
where we have used the definition of $H_{\rm v}(t)$ in Eq.~(\ref{eq:H_theta}), with $\theta(\bar{\bm{r}}) = 2\pi \bar{x}/L_x$ as discussed above Eq.~(\ref{eq: floquet eigenstates with tv on cylinder}), and suppressed the spatial indices $\bm{r}$ and $\bm{r}'$ for notational simplicity, invoking a matrix notation, e.g., $H_{{\rm v}}(t)\chi_{nk}(t)\equiv\sum_{\bm{r}'}H_{{\rm v}}(\bm{r},\bm{r}',t)\chi_{nk}(\bm{r}',t)$. 

Our aim is to show that the right hand side of Eq.~\eqref{eq:time dependent schrodinger} is of the order of $\tfrac{\ell}{L_x}$. 
Performing a position-dependent shift of the time variable in Eq.~\eqref{eq:time dependent schrodinger}, $t\mapsto t + \tfrac{x}{L_x}T$, we find:
\begin{align}
    &\left[i\partial_{t}-H_{{\rm 0}}\left(t-\frac{x'-x}{2L_{x}}T\right)\right]\psi_{nk}(t-\frac{x'-x}{L_{x}}T;\alpha) \nonumber\\
    & =  \left[i\partial_{t}-H_{{\rm 0}}\left(t-\frac{x'-x}{L_{x}}T\right)\right]\psi_{nk}(t-\frac{x'-x}{L_{x}}T;\alpha) \nonumber\\
    &-\frac{x'-x}{2L_{x}}T\frac{\partial H_{0}\left(t-\frac{x'-x}{L_{x}}T\right)}{\partial t}\psi_{nk}(t-\frac{x'-x}{L_{x}}T;\alpha)+O\left(\frac{\ell^{2}}{L_{x}^{2}}\right) \nonumber\\
    &=O\left(\frac{\ell}{L_{x}}\right),
\end{align}
To obtain the second line we expanded $H_0$ around $t-\frac{x'-x}{L_x}T$, and used $|x'-x|<\ell$, where $\ell$ is the range of the hopping in $H_0$.
In obtaining the final expression, we used the Schr\"{o}dinger equation for $\psi_{nk}(t)$ to eliminate the first term in the second line.
(Notice that for any fixed, continuous driving profile, the first-order term in the Taylor expansion can be made dominant over the higher order terms by choosing the circumference of the cylinder, $L_x$, large enough.)

In addition to solving the Schr\"{o}dinger equation, using Eq. \eqref{eq:FloquetThm} and Eq. \eqref{eq: floquet eigenstates with tv on cylinder}, we find that $\chi_{nk}(t)$ satisfies the Floquet boundary condition in time: $\chi_{nk}(t + T) = e^{-i\varepsilon_{nk}(\alpha) T} \chi_{nk}(t)$, where $\alpha$ is determined as described below Eq. (\ref{eq: chi}). This shows that $\chi_{nk}(t)$ is a Floquet eigenstate with quasi-energy $\varepsilon_{nk}(\alpha)$.

We note that, while $\chi_{nk}(\bm{r},t)$ is not a Bloch state, it is an eigenstate of a combined space-time translation:
\begin{align}
\chi_{nk}(\bm{r}+a_0\hat{\bm{x}}, t+\frac{a_0}{L_x}T)=e^{ika_0}\chi_{nk}(\bm{r}, t),
\end{align}
which is a symmetry of the Hamiltonian in the presence of a time vortex.

\subsection{Phase diagram of the Floquet Kitaev honeycomb model}
\label{app: phase diagram}

\begin{figure}
\begin{centering}
\includegraphics[trim={0cm 0cm 0cm 0cm},clip,width=(\textwidth-\columnsep)/2]{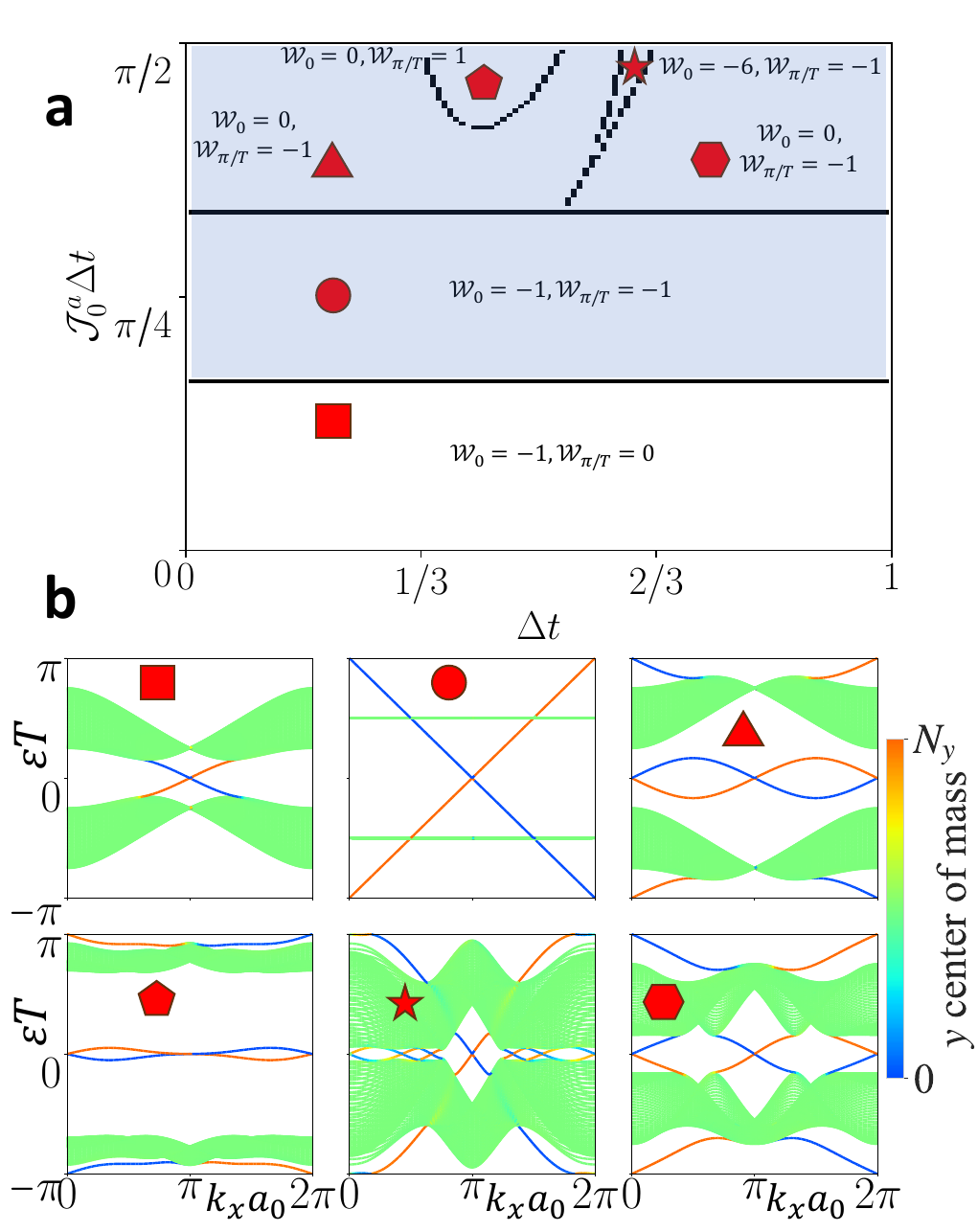}
\par\end{centering}
\caption{\textbf{The phase diagram of the Floquet Kitaev honeycomb model.} \textbf{(a)} The phase diagram parametrized by the pulse duration $\Delta t$ and the pulse integral $\mathcal{J}^a_0\Delta t$. Each phase is labeled by its topological invariants $\mathcal{W}_0, \mathcal{W}_{\pi/T}$. Within the shaded region, the number of chiral edge states through quasi-energy $\pi/T$ is odd and therefore a time vortex defect binds a $\pi$ mode. \textbf{(b)} Band structures of the system with periodic boundary conditions along the $x$ axis and open boundary conditions along the $y$ axis. The different panels show the band structures computed for values of $\Delta t$ and $\mathcal{J}^a_0\Delta t$ marked by red symbols in \textbf{(a)}.}
\label{fig: phase diagram}
\end{figure}

In this section, we map out the phase diagram of the Floquet Kitaev honeycomb model [Eq. \eqref{eq: KSL Hamiltonian in terms of fermions}] in the flux-free sector. This is done by searching the space of parameters (pulse duration $\Delta t$ and the pulse integral $\mathcal{J}^a_0\Delta t$) for gap closings at quasienergy $0$ and $\pi/T$, as  done in Ref.~\cite{Fulga2019}. 
The topological invariants $\mathcal{W}_0$, $\mathcal{W}_{\pi/T}$ are then determined by computing the spectrum on a cylindrical geometry, and counting the number of chiral edge states at each quasi-energy gap, weighted by their chirality. 
The phase diagram is shown in Fig.~\ref{fig: phase diagram}, where the gap closings are indicated by black lines. The resonant point analyzed in the main text is located at $\mathcal{J}^a_0\Delta t=\pi/4$ with non-overlapping pulses ($\Delta t<1/3$) (the red circle belongs to this region).

In the shaded region in Fig.~\ref{fig: phase diagram} (that includes the resonant point)
we find that $\mathcal{W}_{\pi/T}=1 \mod 2$, except on the phase boundaries where $\mathcal{W}_{\pi/T}$ is not defined. Therefore, within this region a time vortex defect binds a Majorana $\pi$ mode, as argued in the main text. In particular, the point examined away from the resonant limit in Fig.~\ref{fig: space-time vortex and spectrum on cylinder}(c) and Fig.~\ref{fig: spectrum} is within the same phase as the resonant point.

\subsection{Kitaev honeycomb model in momentum space}
\label{app: kitaev model in k space}

The $\mathbb{Z}_2$ flux-free sector $W_{p}=1$ can be described by choosing a gauge in which $u_{\mathbf{I},\mathbf{J}}=1$ for $\mathbf{I}$ in the $A$ sublattice and $\mathbf{J}$ in the $B$ sublattice. Here, taking spatially uniform couplings (i.e., in the absence of a time vortex), translation symmetry allows us to write the Hamiltonian \eqref{eq: KSL Hamiltonian in terms of fermions} in momentum space as
\begin{align}
\mathcal{H}=\sum_{\bm{k}}\begin{pmatrix}
c_{-\bm{k},A} & c_{-\bm{k},B}
\end{pmatrix}
H_{\bm{k}}
\begin{pmatrix}
c_{\bm{k},A} \\ c_{\bm{k},B}
\end{pmatrix},
\label{eq: momentum space sum H_k}
\end{align}
where $c_{\bm{k},s}$ is the Fourier transform of $c_{\mathbf{I}}$,
\begin{align}
    c_{\bm{k},s} = \frac{1}{\sqrt{2N_xN_y}}\sum_{i,j}e^{-i\left(k_xai+k_yaj\right)}c_{(i,j,s)}.
\end{align}
Here, for convenience, we take the components $k_x,k_y$ as the projections of the momentum $\bm{k}$ on the (non-orthogonal) primitive vectors of the honeycomb lattice.
$H_{\bm{k}}(t)$ is given by 
\begin{align}
H_{\bm{k}}(t) &= \left[\tilde{\mathcal{J}}^{x}(t)+\tilde{\mathcal{J}}^{y}(t)\cos(k_xa_0)+\tilde{\mathcal{J}}^{z}(t)\cos(k_ya_0)\right]\tau^y\nonumber\\
&+ \left[\tilde{\mathcal{J}}^{y}(t)\sin(k_xa_0)+\tilde{\mathcal{J}}^{z}(t)\sin(k_ya_0)\right]\tau^x,
\label{eq: Hamiltonian in k space}
\end{align}
where the Pauli matrices $\tau^a$ act on the sublattice degree of freedom. We resolve the time-evolution operator in momentum space as $U(\bm{k},t)$ and find that particle-hole symmetry acts as
\begin{align}
    U(\bm{k},t)=U^*(-\bm{k},t).
\end{align}
The time evolution operator can be expressed in terms of a spectral decomposition
\begin{align}
    U(\bm{k},t)=\sum_{n}\left|\psi_n(\bm{k},t)\right\rangle\left\langle\psi_n(\bm{k},t)\right|e^{-i\phi_n(\bm{k},t)},
\end{align}
where $\phi_n(\bm{k},t)$ are called the phase bands \cite{Nathan2015}, and are defined modulo $2\pi$. At $t=T$, $\varepsilon_n(\bm{k})=\phi_n(\bm{k},T)/T$ is referred to as the quasienergy, and unlike the bands of a static system, these are periodic with period $\frac{2\pi}{T}$. Particle-hole symmetry maps $\phi_n(\bm{k},t)$ to $-\phi_n(-\bm{k},t)$, such that there are two particle-hole symmetric quasienergies, $\varepsilon=0$ and $\tfrac{\pi}{T}$, at which protected Majorana modes can appear.

\subsection{Computation of the topological invariants}
\label{app: Topological invariants resolved on time-like surfaces}

This section provides the details of the numerical calculation yielding the topological invariants of the time vortex defect in the driven Kitaev honeycomb model. We specialize to the resonant limit discussed in Sec.~\ref{sec: resonant limit} and follow the procedure outlined in Sec.~\ref{sec: invariants}.
We set the duration of the pulses during which $\tilde{\mathcal{J}}^a$ are non-zero to be $\Delta t = T/3$. Since the invariant only requires knowledge of the time evolution of a translationally invariant system, the pulses corresponding to different couplings do not overlap.

We begin by numerically mapping out the locations of the ``topological singularities'' 
in the three dimensional space of $k_y,\lambda,t$ for both $k_x=0$ and $\tfrac{\pi}{a_0}$ by exponentiating the Hamiltonian \eqref{eq: Hamiltonian in k space} for each pulse separately and taking products of these unitaries.
For $k_x=0$, we find that the $\phi=0$ singularities are located on a one-dimensional curve lying in the $k_y=\tfrac{\pi}{a_0}$ plane, originating at $t=0$ and extending up to $t=\frac{2}{3}T$. On this plane, $H_{\bm{k}} = \left(\tilde{\mathcal{J}}^{x}+\tilde{\mathcal{J}}^{y}-\tilde{\mathcal{J}}^{z}\right)\tau^y$. Therefore, time evolution on this plane corresponds to rotation around the $\tau^y$ axis of the Bloch sphere in the positive direction during the action of the $x$ and $y$ pulses, and in the negative direction during the action of the $z$ pulse. The topological singularities at $\phi=0$ correspond to the cancellation of the $z$ pulse with either the preceding $y$ pulse or the successive $x$ pulse, or to a cancellation of complementary parts of $x$ and $y$ pulses with the $z$ pulse acting in between. Singularities at $\phi=\pi$ are found on the plane $t=\frac{2}{3}T$, both on the line at $k_y=0$ and on the line at $\lambda=0$. At $k_x=k_y=0$ all pulses act identically, so for any $\lambda$, two thirds of a driving period leads to the accumulation of two pulses. At $\lambda=0$ the singularity is due to the accumulation of an $x$ and a $y$ pulse which are equivalent at $k_x=0$ for all $k_y$.

At $k_x=\tfrac{\pi}{a_0}$, the $\phi=0$ singularities are found both on 1D curves and on a 2D plane. These are all smoothly connected to the plane $t=0$ and extend up to $t=\tfrac{2}{3}T$. 
On this plane there is a cancellation of 
the $x$ and $y$ pulses. Since during this part of the period $\tilde{\mathcal{J}}^{z}=0$, the plane is parallel to the $k_y$ axis. The line segment singularities appear on the planes $k_y=0,\tfrac{\pi}{a_0}$ and correspond to the cancellation of the $z$ pulse with the $y$ or $x$ pulse respectively. The singularities at $\phi=\pi$ are again confined to the $t=\tfrac{2}{3}T$ plane, but here they appear at two isolated points at $k_y=0,\lambda=\tfrac{1}{3}$, and $k_y=\tfrac{\pi}{a_0},\lambda=\tfrac{2}{3}$, and are due to the accumulation of $z$ and $x$ or $y$ and $z$ pulses respectively.

\begin{figure}
\begin{centering}
\includegraphics[trim={0cm 0cm 0cm 0cm},clip,width=(\textwidth-\columnsep)/2]{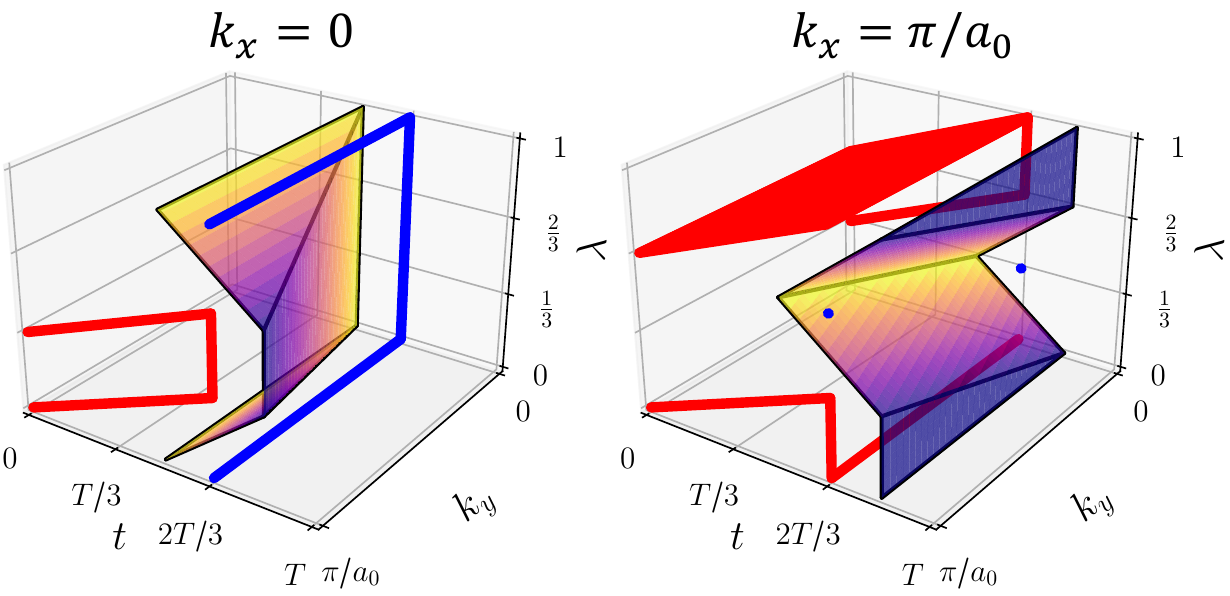}
\par\end{centering}
\caption{\textbf{Topological singularities of the phase bands.} The topological singularities at $\phi=0$ (red) and $\phi=\pi$ (blue) of the effective 1D model at the particle-hole symmetric points $k_x=0,\pi/a_0$. For an interactive version, see \cite{link_interactive}. The topological invariant of an effective non-driven model is computed on two time-like surfaces in the $k_y,\theta,t$ space for each $k_x$. The first surface (shown) is chosen to be piece-wise linear such that it separates the two types of singularities and is periodic in $\lambda$ with period $1$ and the second surface taken at $t=T$.}
\label{fig: real topological singularities}
\end{figure}

Next, for $k_x=0$ and $k_x=\tfrac{\pi}{a_0}$ separately, we construct surfaces which are periodic in $\lambda$ and separate the two types of singularities. Since both types of singularities exist on the plane $t=\frac{2}{3}T$, this cannot be accomplished by planes of constant $t$. Instead, we design suitable piece-wise planar surfaces composed of edge sharing polygons, as shown along with the singularities in Fig.~\ref{fig: real topological singularities}.

Then, as explained in the main text, we compute a $\mathbb{Z}_2$ topological invariant on each one of these surfaces, and on the $t=T$ plane. On each surface, we define the effective gapped Hamiltonian $H_{\rm{eff}}(\bm{k},t;2\pi\lambda)$ [Eq.~\eqref{eq:H_eff}]. We describe the procedure for computing the invariant on a plane of constant $t$; the generalization to any surface characterized by a function $t=f(k_y,\lambda)$ which is periodic in $\lambda$ is straightforward.

The $\mathbb{Z}_2$ invariant characterizing the surface is the fermion parity pump invariant~\cite{Teo2010} of the effective one-dimensional static Hamiltonian defined by $H_{\rm{eff}}$ as a function of $k_y$, when the parameter $\lambda$ changes adiabatically from $0$ to $1$. To compute it, we follow the procedure described in Ref. \cite{Keselman2013}. We first diagonalize the Hamiltonian $H_{\rm{eff}}$ for each $k_y$ and $\lambda$, denoting the diagonalizing matrix by $S(k_y,\lambda)$, where each row of $S$ is an eigenvector of $H_{\rm{eff}}$. Importantly, it is possible to find a continuous gauge choice for the rows of $S(k_y,\lambda)$ in the domain $k_y\in 
[0,\tfrac{\pi}{a_0}]$, $\lambda\in [0,1)$, because $S(k_y,\lambda)$ is not required to be periodic in $k_y$ within this domain (which spans only part of the Brillouin zone).

For $k_x$ and $k_y$ both equal to $0$ or $\tfrac{\pi}{a_0}$, $H_{\rm{eff}}(\bm{k},t;2\pi\lambda)$ is particle-hole symmetric, and can be chosen to be purely imaginary, $H_{\rm{eff}}=-H^{*}_{\rm{eff}}$ (as is manifestly the case for the driven Kitaev model, Eq.~\ref{eq: Hamiltonian in k space}). 
Then, the eigenvectors of $H_{\rm{eff}}$ come in pairs, $\{\vert u \rangle, \vert u^{*} \rangle\}$, with eigenvalues $\{\varepsilon,-\varepsilon\}$, where $ \vert u^{*} \rangle$ is the complex conjugate of $\vert u \rangle$. We define a matrix $O(k_y\in\{0,\tfrac{\pi}{a_0}\},\lambda)$ whose rows are formed of the real and imaginary parts of the rows of $S(k_y\in\{0,\tfrac{\pi}{a_0}\},\lambda)$: the corresponding rows of $O$ are $\{\tfrac{1}{\sqrt{2}}(\vert u\rangle+\vert u^*\rangle), \tfrac{1}{i\sqrt{2}}(\vert u\rangle-\vert u^*\rangle)\}$. By the hermiticity of $H_{\rm{eff}}$, $\vert u\rangle$ and $\vert u^*\rangle$ are orthogonal, and hence $O$ is an orthogonal matrix.
Since $\pi_1[SO(N)]=\mathbb{Z}_2$, two $\mathbb{Z}_2$ invariants can be defined per $k_x\in\{0,\tfrac{\pi}{a_0}\}$, corresponding to $O(k_x,k_y=0,\lambda)$ and $O(k_x,k_y=\tfrac{\pi}{a_0},\lambda)$. 

To compute the invariants in practice, we note that as explained in Ref.~\cite{Keselman2013}, the $\mathbb{Z}_2$ invariants are equal to the parity of the number of times the phases of the eigenvalues of $O(k_x,k_y\in\{0,\tfrac{\pi}{a_0}\},\lambda)$ cross $\pi$ modulo $2\pi$ as $\lambda$ varies from $0$ to $1$. The invariants corresponding to $k_y=0$ and $\tfrac{\pi}{a_0}$ depend on the gauge choice, but their sum (mod 2) is gauge invariant, because of the requirement that $S(k_x,k_y,\lambda)$ is continuous in the entire domain $k_y\in [0,\tfrac{\pi}{a_0}]$, and is equal to the $\mathbb{Z}_2$ invariant characterizing $H_{\rm{eff}}$ on the surface $t=f(k_y,\lambda)$. The values of the invariants on the surfaces separating the phase singularities and on $t=T$ are shown in Table \ref{tab: static system invariants}. 

\begin{table}[]
    \centering
    \begin{tabular}{c|c|c|c}
         & $t\stackrel{*}{=}\frac{2}{3}T$ & $t=T$ \\
         & after $\phi=0$ singularities & after $\phi=\pi$ singularities\\
         \hline
         $k_x=0$ & $0$ & $1$ \\
         $k_x=\tfrac{\pi}{a_0}$ & $0$ & $0$
    \end{tabular}
    \caption{The $\mathbb{Z}_2$ topological invariants of the effective non-driven 1D systems defined on time-like surfaces separating the topological singularities in the phase bands of the 1D driven model for the time vortex defect. The surfaces at $t=\frac{2}{3}T$ are taken to be piece-wise linear in order to avoid intersecting the singularities, as shown in Fig.~\ref{fig: real topological singularities}.}
    \label{tab: static system invariants}
\end{table}

Finally, as discussed in the main text, we identify $\nu(\varepsilon=0,k_x) + \nu(\varepsilon=\tfrac{\pi}{T},k_x) \pmod{2}$ as the two $\mathbb{Z}_2$ invariants computed on the $t=T$ planes. Adding the two invariants for $k_x=0$ and $k_x=\tfrac{\pi}{a_0}$ gives $\nu_0 + \nu_\pi \pmod{2}$. To disentangle $\nu_0$ and $\nu_\pi$, we subtract (mod 2) the invariant on the $t=T$ plane from the invariant of the surface that separates the $\phi=\pi$ phase singularities from the $\phi=0$ phase singularities. For the Kitaev model example, these surfaces are shown in Fig.~\ref{fig: real topological singularities} for $k_x=0$ and $\tfrac{\pi}{a_0}$. As explained in the main text, the result is the change in the $\mathbb{Z}_2$ invariant that corresponds to the $\phi=\pi$ gap, since the invariant can only change by a phase singularity (gap closing). The change in the invariant across the phase singularities gives $\nu_\pi$, because the $\phi=\pi$ invariant is trivial on the plane $t=0$. We thus have both $\nu_0 + \nu_\pi \pmod{2}$ and $\nu_\pi$, from which we can deduce also $\nu_0$. Using this procedure for the time vortex in the resonantly driven Kitaev model, we see from Table \ref{tab: static system invariants} that $\nu_0=0$, $\nu_\pi=1$, as expected.

\section{Data availability}
The data that support the findings of this study are available at \url{https://github.com/kishonyWIS/free_fermions/tree/FloquetKSL}.

\section{Code availability}
The code used in this work is available at \url{https://github.com/kishonyWIS/free_fermions/tree/FloquetKSL}.

\acknowledgements

We thank Ady Stern for useful discussions. Work at the Weizmann Institute was supported by CRC 183 of the Deutsche Forschungsgemeinschaft (project number 277101999, subproject A01), and a research grant from the Estate of Gerald Alexander. N.L. and E.B. acknowledge support from ISF-MAFAT Quantum Science and Technology Grant no. 2478/24. M.R. acknowledges the Brown Institute for Basic Sciences, administered by Caltech and established by Ross M. Brown, the University of Washington College of Arts and Sciences, and the Kenneth K. Young Memorial Professorship for support.

\section{Author Contributions}

G.K., O.G., N.H.L., M.S.R., and E.B.~conceived the project, performed the theoretical analysis, and wrote the paper. G.K.~performed the numerical simulations.

\section{Competing interests}
The authors declare no competing interests.

\bibliography{literature.bib,floquetMajoranaLiterature.bib}

\begin{thebibliography}{41}%
\makeatletter
\providecommand \@ifxundefined [1]{%
 \@ifx{#1\undefined}
}%
\providecommand \@ifnum [1]{%
 \ifnum #1\expandafter \@firstoftwo
 \else \expandafter \@secondoftwo
 \fi
}%
\providecommand \@ifx [1]{%
 \ifx #1\expandafter \@firstoftwo
 \else \expandafter \@secondoftwo
 \fi
}%
\providecommand \natexlab [1]{#1}%
\providecommand \enquote  [1]{``#1''}%
\providecommand \bibnamefont  [1]{#1}%
\providecommand \bibfnamefont [1]{#1}%
\providecommand \citenamefont [1]{#1}%
\providecommand \href@noop [0]{\@secondoftwo}%
\providecommand \href [0]{\begingroup \@sanitize@url \@href}%
\providecommand \@href[1]{\@@startlink{#1}\@@href}%
\providecommand \@@href[1]{\endgroup#1\@@endlink}%
\providecommand \@sanitize@url [0]{\catcode `\\12\catcode `\$12\catcode
  `\&12\catcode `\#12\catcode `\^12\catcode `\_12\catcode `\%12\relax}%
\providecommand \@@startlink[1]{}%
\providecommand \@@endlink[0]{}%
\providecommand \url  [0]{\begingroup\@sanitize@url \@url }%
\providecommand \@url [1]{\endgroup\@href {#1}{\urlprefix }}%
\providecommand \urlprefix  [0]{URL }%
\providecommand \Eprint [0]{\href }%
\providecommand \doibase [0]{http://dx.doi.org/}%
\providecommand \selectlanguage [0]{\@gobble}%
\providecommand \bibinfo  [0]{\@secondoftwo}%
\providecommand \bibfield  [0]{\@secondoftwo}%
\providecommand \translation [1]{[#1]}%
\providecommand \BibitemOpen [0]{}%
\providecommand \bibitemStop [0]{}%
\providecommand \bibitemNoStop [0]{.\EOS\space}%
\providecommand \EOS [0]{\spacefactor3000\relax}%
\providecommand \BibitemShut  [1]{\csname bibitem#1\endcsname}%
\let\auto@bib@innerbib\@empty
\bibitem [{\citenamefont {Kitagawa}\ \emph {et~al.}(2010)\citenamefont
  {Kitagawa}, \citenamefont {Berg}, \citenamefont {Rudner},\ and\ \citenamefont
  {Demler}}]{Kitagawa2010}%
  \BibitemOpen
  \bibfield  {author} {\bibinfo {author} {\bibfnamefont {Takuya}\ \bibnamefont
  {Kitagawa}}, \bibinfo {author} {\bibfnamefont {Erez}\ \bibnamefont {Berg}},
  \bibinfo {author} {\bibfnamefont {Mark}\ \bibnamefont {Rudner}}, \ and\
  \bibinfo {author} {\bibfnamefont {Eugene}\ \bibnamefont {Demler}},\
  }\bibfield  {title} {\enquote {\bibinfo {title} {Topological characterization
  of periodically driven quantum systems},}\ }\href {\doibase
  10.1103/PhysRevB.82.235114} {\bibfield  {journal} {\bibinfo  {journal} {Phys.
  Rev. B}\ }\textbf {\bibinfo {volume} {82}},\ \bibinfo {pages} {235114}
  (\bibinfo {year} {2010})}\BibitemShut {NoStop}%
\bibitem [{\citenamefont {Jiang}\ \emph {et~al.}(2011)\citenamefont {Jiang},
  \citenamefont {Kitagawa}, \citenamefont {Alicea}, \citenamefont {Akhmerov},
  \citenamefont {Pekker}, \citenamefont {Refael}, \citenamefont {Cirac},
  \citenamefont {Demler}, \citenamefont {Lukin},\ and\ \citenamefont
  {Zoller}}]{Jiang2011}%
  \BibitemOpen
  \bibfield  {author} {\bibinfo {author} {\bibfnamefont {Liang}\ \bibnamefont
  {Jiang}}, \bibinfo {author} {\bibfnamefont {Takuya}\ \bibnamefont
  {Kitagawa}}, \bibinfo {author} {\bibfnamefont {Jason}\ \bibnamefont
  {Alicea}}, \bibinfo {author} {\bibfnamefont {A.~R.}\ \bibnamefont
  {Akhmerov}}, \bibinfo {author} {\bibfnamefont {David}\ \bibnamefont
  {Pekker}}, \bibinfo {author} {\bibfnamefont {Gil}\ \bibnamefont {Refael}},
  \bibinfo {author} {\bibfnamefont {J.~Ignacio}\ \bibnamefont {Cirac}},
  \bibinfo {author} {\bibfnamefont {Eugene}\ \bibnamefont {Demler}}, \bibinfo
  {author} {\bibfnamefont {Mikhail~D.}\ \bibnamefont {Lukin}}, \ and\ \bibinfo
  {author} {\bibfnamefont {Peter}\ \bibnamefont {Zoller}},\ }\bibfield  {title}
  {\enquote {\bibinfo {title} {{Majorana Fermions in Equilibrium and in Driven
  Cold-Atom Quantum Wires}},}\ }\href {\doibase 10.1103/physrevlett.106.220402}
  {\bibfield  {journal} {\bibinfo  {journal} {Physical Review Letters}\
  }\textbf {\bibinfo {volume} {106}} (\bibinfo {year} {2011}),\
  10.1103/physrevlett.106.220402}\BibitemShut {NoStop}%
\bibitem [{\citenamefont {Rudner}\ \emph {et~al.}(2013)\citenamefont {Rudner},
  \citenamefont {Lindner}, \citenamefont {Berg},\ and\ \citenamefont
  {Levin}}]{Rudner2013}%
  \BibitemOpen
  \bibfield  {author} {\bibinfo {author} {\bibfnamefont {Mark~S.}\ \bibnamefont
  {Rudner}}, \bibinfo {author} {\bibfnamefont {Netanel~H.}\ \bibnamefont
  {Lindner}}, \bibinfo {author} {\bibfnamefont {Erez}\ \bibnamefont {Berg}}, \
  and\ \bibinfo {author} {\bibfnamefont {Michael}\ \bibnamefont {Levin}},\
  }\bibfield  {title} {\enquote {\bibinfo {title} {Anomalous edge states and
  the bulk-edge correspondence for periodically driven two-dimensional
  systems},}\ }\href {\doibase 10.1103/PhysRevX.3.031005} {\bibfield  {journal}
  {\bibinfo  {journal} {Phys. Rev. X}\ }\textbf {\bibinfo {volume} {3}},\
  \bibinfo {pages} {031005} (\bibinfo {year} {2013})}\BibitemShut {NoStop}%
\bibitem [{\citenamefont {Nathan}\ and\ \citenamefont
  {Rudner}(2015)}]{Nathan2015}%
  \BibitemOpen
  \bibfield  {author} {\bibinfo {author} {\bibfnamefont {Frederik}\
  \bibnamefont {Nathan}}\ and\ \bibinfo {author} {\bibfnamefont {Mark~S.}\
  \bibnamefont {Rudner}},\ }\bibfield  {title} {\enquote {\bibinfo {title}
  {{Topological singularities and the general classification of Floquet–Bloch
  systems}},}\ }\href {\doibase 10.1088/1367-2630/17/12/125014} {\bibfield
  {journal} {\bibinfo  {journal} {New Journal of Physics}\ }\textbf {\bibinfo
  {volume} {17}},\ \bibinfo {pages} {125014} (\bibinfo {year}
  {2015})}\BibitemShut {NoStop}%
\bibitem [{\citenamefont {Titum}\ \emph {et~al.}(2016)\citenamefont {Titum},
  \citenamefont {Berg}, \citenamefont {Rudner}, \citenamefont {Refael},\ and\
  \citenamefont {Lindner}}]{Titum2016}%
  \BibitemOpen
  \bibfield  {author} {\bibinfo {author} {\bibfnamefont {Paraj}\ \bibnamefont
  {Titum}}, \bibinfo {author} {\bibfnamefont {Erez}\ \bibnamefont {Berg}},
  \bibinfo {author} {\bibfnamefont {Mark~S.}\ \bibnamefont {Rudner}}, \bibinfo
  {author} {\bibfnamefont {Gil}\ \bibnamefont {Refael}}, \ and\ \bibinfo
  {author} {\bibfnamefont {Netanel~H.}\ \bibnamefont {Lindner}},\ }\bibfield
  {title} {\enquote {\bibinfo {title} {Anomalous floquet-anderson insulator as
  a nonadiabatic quantized charge pump},}\ }\href {\doibase
  10.1103/PhysRevX.6.021013} {\bibfield  {journal} {\bibinfo  {journal} {Phys.
  Rev. X}\ }\textbf {\bibinfo {volume} {6}},\ \bibinfo {pages} {021013}
  (\bibinfo {year} {2016})}\BibitemShut {NoStop}%
\bibitem [{\citenamefont {von Keyserlingk}\ and\ \citenamefont
  {Sondhi}(2016)}]{Keyserlingk2016}%
  \BibitemOpen
  \bibfield  {author} {\bibinfo {author} {\bibfnamefont {C.~W.}\ \bibnamefont
  {von Keyserlingk}}\ and\ \bibinfo {author} {\bibfnamefont {S.~L.}\
  \bibnamefont {Sondhi}},\ }\bibfield  {title} {\enquote {\bibinfo {title}
  {Phase structure of one-dimensional interacting floquet systems. i. abelian
  symmetry-protected topological phases},}\ }\href {\doibase
  10.1103/PhysRevB.93.245145} {\bibfield  {journal} {\bibinfo  {journal} {Phys.
  Rev. B}\ }\textbf {\bibinfo {volume} {93}},\ \bibinfo {pages} {245145}
  (\bibinfo {year} {2016})}\BibitemShut {NoStop}%
\bibitem [{\citenamefont {Roy}\ and\ \citenamefont {Harper}(2016)}]{Roy2016}%
  \BibitemOpen
  \bibfield  {author} {\bibinfo {author} {\bibfnamefont {Rahul}\ \bibnamefont
  {Roy}}\ and\ \bibinfo {author} {\bibfnamefont {Fenner}\ \bibnamefont
  {Harper}},\ }\bibfield  {title} {\enquote {\bibinfo {title} {Abelian floquet
  symmetry-protected topological phases in one dimension},}\ }\href {\doibase
  10.1103/PhysRevB.94.125105} {\bibfield  {journal} {\bibinfo  {journal} {Phys.
  Rev. B}\ }\textbf {\bibinfo {volume} {94}},\ \bibinfo {pages} {125105}
  (\bibinfo {year} {2016})}\BibitemShut {NoStop}%
\bibitem [{\citenamefont {Khemani}\ \emph {et~al.}(2016)\citenamefont
  {Khemani}, \citenamefont {Lazarides}, \citenamefont {Moessner},\ and\
  \citenamefont {Sondhi}}]{Khemani2016}%
  \BibitemOpen
  \bibfield  {author} {\bibinfo {author} {\bibfnamefont {Vedika}\ \bibnamefont
  {Khemani}}, \bibinfo {author} {\bibfnamefont {Achilleas}\ \bibnamefont
  {Lazarides}}, \bibinfo {author} {\bibfnamefont {Roderich}\ \bibnamefont
  {Moessner}}, \ and\ \bibinfo {author} {\bibfnamefont {S.~L.}\ \bibnamefont
  {Sondhi}},\ }\bibfield  {title} {\enquote {\bibinfo {title} {Phase structure
  of driven quantum systems},}\ }\href {\doibase
  10.1103/PhysRevLett.116.250401} {\bibfield  {journal} {\bibinfo  {journal}
  {Phys. Rev. Lett.}\ }\textbf {\bibinfo {volume} {116}},\ \bibinfo {pages}
  {250401} (\bibinfo {year} {2016})}\BibitemShut {NoStop}%
\bibitem [{\citenamefont {Else}\ and\ \citenamefont {Nayak}(2016)}]{Else2016}%
  \BibitemOpen
  \bibfield  {author} {\bibinfo {author} {\bibfnamefont {Dominic~V.}\
  \bibnamefont {Else}}\ and\ \bibinfo {author} {\bibfnamefont {Chetan}\
  \bibnamefont {Nayak}},\ }\bibfield  {title} {\enquote {\bibinfo {title}
  {Classification of topological phases in periodically driven interacting
  systems},}\ }\href {\doibase 10.1103/PhysRevB.93.201103} {\bibfield
  {journal} {\bibinfo  {journal} {Phys. Rev. B}\ }\textbf {\bibinfo {volume}
  {93}},\ \bibinfo {pages} {201103} (\bibinfo {year} {2016})}\BibitemShut
  {NoStop}%
\bibitem [{\citenamefont {Roy}\ and\ \citenamefont {Harper}(2017)}]{Roy2017}%
  \BibitemOpen
  \bibfield  {author} {\bibinfo {author} {\bibfnamefont {Rahul}\ \bibnamefont
  {Roy}}\ and\ \bibinfo {author} {\bibfnamefont {Fenner}\ \bibnamefont
  {Harper}},\ }\bibfield  {title} {\enquote {\bibinfo {title} {Periodic table
  for floquet topological insulators},}\ }\href {\doibase
  10.1103/PhysRevB.96.155118} {\bibfield  {journal} {\bibinfo  {journal} {Phys.
  Rev. B}\ }\textbf {\bibinfo {volume} {96}},\ \bibinfo {pages} {155118}
  (\bibinfo {year} {2017})}\BibitemShut {NoStop}%
\bibitem [{\citenamefont {Oka}\ and\ \citenamefont
  {Kitamura}(2019)}]{oka2019floquet}%
  \BibitemOpen
  \bibfield  {author} {\bibinfo {author} {\bibfnamefont {Takashi}\ \bibnamefont
  {Oka}}\ and\ \bibinfo {author} {\bibfnamefont {Sota}\ \bibnamefont
  {Kitamura}},\ }\bibfield  {title} {\enquote {\bibinfo {title} {Floquet
  engineering of quantum materials},}\ }\href
  {https://www.annualreviews.org/content/journals/10.1146/annurev-conmatphys-031218-013423}
  {\bibfield  {journal} {\bibinfo  {journal} {Annual Review of Condensed Matter
  Physics}\ }\textbf {\bibinfo {volume} {10}},\ \bibinfo {pages} {387--408}
  (\bibinfo {year} {2019})}\BibitemShut {NoStop}%
\bibitem [{\citenamefont {Rudner}\ and\ \citenamefont
  {Lindner}(2020)}]{rudner2020floquet}%
  \BibitemOpen
  \bibfield  {author} {\bibinfo {author} {\bibfnamefont {Mark~S.}\ \bibnamefont
  {Rudner}}\ and\ \bibinfo {author} {\bibfnamefont {Netanel~H.}\ \bibnamefont
  {Lindner}},\ }\href@noop {} {\enquote {\bibinfo {title} {The floquet
  engineer's handbook},}\ } (\bibinfo {year} {2020}),\ \Eprint
  {http://arxiv.org/abs/2003.08252} {arXiv:2003.08252 [cond-mat.mes-hall]}
  \BibitemShut {NoStop}%
\bibitem [{\citenamefont {Kitagawa}\ \emph {et~al.}(2012)\citenamefont
  {Kitagawa}, \citenamefont {Broome}, \citenamefont {Fedrizzi}, \citenamefont
  {Rudner}, \citenamefont {Berg}, \citenamefont {Kassal}, \citenamefont
  {Aspuru-Guzik}, \citenamefont {Demler},\ and\ \citenamefont
  {White}}]{kitagawa2012observation}%
  \BibitemOpen
  \bibfield  {author} {\bibinfo {author} {\bibfnamefont {Takuya}\ \bibnamefont
  {Kitagawa}}, \bibinfo {author} {\bibfnamefont {Matthew~A}\ \bibnamefont
  {Broome}}, \bibinfo {author} {\bibfnamefont {Alessandro}\ \bibnamefont
  {Fedrizzi}}, \bibinfo {author} {\bibfnamefont {Mark~S}\ \bibnamefont
  {Rudner}}, \bibinfo {author} {\bibfnamefont {Erez}\ \bibnamefont {Berg}},
  \bibinfo {author} {\bibfnamefont {Ivan}\ \bibnamefont {Kassal}}, \bibinfo
  {author} {\bibfnamefont {Al{\'a}n}\ \bibnamefont {Aspuru-Guzik}}, \bibinfo
  {author} {\bibfnamefont {Eugene}\ \bibnamefont {Demler}}, \ and\ \bibinfo
  {author} {\bibfnamefont {Andrew~G}\ \bibnamefont {White}},\ }\bibfield
  {title} {\enquote {\bibinfo {title} {Observation of topologically protected
  bound states in photonic quantum walks},}\ }\href
  {https://www.nature.com/articles/ncomms1872} {\bibfield  {journal} {\bibinfo
  {journal} {Nature communications}\ }\textbf {\bibinfo {volume} {3}},\
  \bibinfo {pages} {882} (\bibinfo {year} {2012})}\BibitemShut {NoStop}%
\bibitem [{\citenamefont {Liu}\ \emph {et~al.}(2013)\citenamefont {Liu},
  \citenamefont {Levchenko},\ and\ \citenamefont {Baranger}}]{Liu2013}%
  \BibitemOpen
  \bibfield  {author} {\bibinfo {author} {\bibfnamefont {Dong~E.}\ \bibnamefont
  {Liu}}, \bibinfo {author} {\bibfnamefont {Alex}\ \bibnamefont {Levchenko}}, \
  and\ \bibinfo {author} {\bibfnamefont {Harold~U.}\ \bibnamefont {Baranger}},\
  }\bibfield  {title} {\enquote {\bibinfo {title} {Floquet majorana fermions
  for topological qubits in superconducting devices and cold-atom systems},}\
  }\href {\doibase 10.1103/PhysRevLett.111.047002} {\bibfield  {journal}
  {\bibinfo  {journal} {Phys. Rev. Lett.}\ }\textbf {\bibinfo {volume} {111}},\
  \bibinfo {pages} {047002} (\bibinfo {year} {2013})}\BibitemShut {NoStop}%
\bibitem [{\citenamefont {Bomantara}\ and\ \citenamefont
  {Gong}(2018)}]{Bomantara2018}%
  \BibitemOpen
  \bibfield  {author} {\bibinfo {author} {\bibfnamefont {Raditya~Weda}\
  \bibnamefont {Bomantara}}\ and\ \bibinfo {author} {\bibfnamefont {Jiangbin}\
  \bibnamefont {Gong}},\ }\bibfield  {title} {\enquote {\bibinfo {title}
  {Simulation of non-abelian braiding in majorana time crystals},}\ }\href
  {\doibase 10.1103/PhysRevLett.120.230405} {\bibfield  {journal} {\bibinfo
  {journal} {Phys. Rev. Lett.}\ }\textbf {\bibinfo {volume} {120}},\ \bibinfo
  {pages} {230405} (\bibinfo {year} {2018})}\BibitemShut {NoStop}%
\bibitem [{\citenamefont {Bauer}\ \emph {et~al.}(2019)\citenamefont {Bauer},
  \citenamefont {Pereg-Barnea}, \citenamefont {Karzig}, \citenamefont {Rieder},
  \citenamefont {Refael}, \citenamefont {Berg},\ and\ \citenamefont
  {Oreg}}]{Bauer2019}%
  \BibitemOpen
  \bibfield  {author} {\bibinfo {author} {\bibfnamefont {Bela}\ \bibnamefont
  {Bauer}}, \bibinfo {author} {\bibfnamefont {T.}~\bibnamefont {Pereg-Barnea}},
  \bibinfo {author} {\bibfnamefont {Torsten}\ \bibnamefont {Karzig}}, \bibinfo
  {author} {\bibfnamefont {Maria-Theresa}\ \bibnamefont {Rieder}}, \bibinfo
  {author} {\bibfnamefont {Gil}\ \bibnamefont {Refael}}, \bibinfo {author}
  {\bibfnamefont {Erez}\ \bibnamefont {Berg}}, \ and\ \bibinfo {author}
  {\bibfnamefont {Yuval}\ \bibnamefont {Oreg}},\ }\bibfield  {title} {\enquote
  {\bibinfo {title} {{Topologically protected braiding in a single wire using
  Floquet Majorana modes}},}\ }\href {\doibase 10.1103/physrevb.100.041102}
  {\bibfield  {journal} {\bibinfo  {journal} {Physical Review B}\ }\textbf
  {\bibinfo {volume} {100}} (\bibinfo {year} {2019}),\
  10.1103/physrevb.100.041102}\BibitemShut {NoStop}%
\bibitem [{\citenamefont {Matthies}\ \emph {et~al.}(2022)\citenamefont
  {Matthies}, \citenamefont {Park}, \citenamefont {Berg},\ and\ \citenamefont
  {Rosch}}]{Matthies2022}%
  \BibitemOpen
  \bibfield  {author} {\bibinfo {author} {\bibfnamefont {Anne}\ \bibnamefont
  {Matthies}}, \bibinfo {author} {\bibfnamefont {Jinhong}\ \bibnamefont
  {Park}}, \bibinfo {author} {\bibfnamefont {Erez}\ \bibnamefont {Berg}}, \
  and\ \bibinfo {author} {\bibfnamefont {Achim}\ \bibnamefont {Rosch}},\
  }\bibfield  {title} {\enquote {\bibinfo {title} {Stability of floquet
  majorana box qubits},}\ }\href {\doibase 10.1103/PhysRevLett.128.127702}
  {\bibfield  {journal} {\bibinfo  {journal} {Phys. Rev. Lett.}\ }\textbf
  {\bibinfo {volume} {128}},\ \bibinfo {pages} {127702} (\bibinfo {year}
  {2022})}\BibitemShut {NoStop}%
\bibitem [{\citenamefont {Wilczek}(2012)}]{Wilczek2012}%
  \BibitemOpen
  \bibfield  {author} {\bibinfo {author} {\bibfnamefont {Frank}\ \bibnamefont
  {Wilczek}},\ }\bibfield  {title} {\enquote {\bibinfo {title} {Quantum time
  crystals},}\ }\href {\doibase 10.1103/PhysRevLett.109.160401} {\bibfield
  {journal} {\bibinfo  {journal} {Phys. Rev. Lett.}\ }\textbf {\bibinfo
  {volume} {109}},\ \bibinfo {pages} {160401} (\bibinfo {year}
  {2012})}\BibitemShut {NoStop}%
\bibitem [{\citenamefont {von Keyserlingk}\ \emph {et~al.}(2016)\citenamefont
  {von Keyserlingk}, \citenamefont {Khemani},\ and\ \citenamefont
  {Sondhi}}]{Khemani2016a}%
  \BibitemOpen
  \bibfield  {author} {\bibinfo {author} {\bibfnamefont {C.~W.}\ \bibnamefont
  {von Keyserlingk}}, \bibinfo {author} {\bibfnamefont {Vedika}\ \bibnamefont
  {Khemani}}, \ and\ \bibinfo {author} {\bibfnamefont {S.~L.}\ \bibnamefont
  {Sondhi}},\ }\bibfield  {title} {\enquote {\bibinfo {title} {Absolute
  stability and spatiotemporal long-range order in floquet systems},}\ }\href
  {\doibase 10.1103/PhysRevB.94.085112} {\bibfield  {journal} {\bibinfo
  {journal} {Phys. Rev. B}\ }\textbf {\bibinfo {volume} {94}},\ \bibinfo
  {pages} {085112} (\bibinfo {year} {2016})}\BibitemShut {NoStop}%
\bibitem [{\citenamefont {Else}\ \emph {et~al.}(2016)\citenamefont {Else},
  \citenamefont {Bauer},\ and\ \citenamefont {Nayak}}]{Else2016a}%
  \BibitemOpen
  \bibfield  {author} {\bibinfo {author} {\bibfnamefont {Dominic~V.}\
  \bibnamefont {Else}}, \bibinfo {author} {\bibfnamefont {Bela}\ \bibnamefont
  {Bauer}}, \ and\ \bibinfo {author} {\bibfnamefont {Chetan}\ \bibnamefont
  {Nayak}},\ }\bibfield  {title} {\enquote {\bibinfo {title} {Floquet time
  crystals},}\ }\href {\doibase 10.1103/PhysRevLett.117.090402} {\bibfield
  {journal} {\bibinfo  {journal} {Phys. Rev. Lett.}\ }\textbf {\bibinfo
  {volume} {117}},\ \bibinfo {pages} {090402} (\bibinfo {year}
  {2016})}\BibitemShut {NoStop}%
\bibitem [{\citenamefont {Morimoto}\ \emph {et~al.}(2017)\citenamefont
  {Morimoto}, \citenamefont {Po},\ and\ \citenamefont
  {Vishwanath}}]{Morimoto2017}%
  \BibitemOpen
  \bibfield  {author} {\bibinfo {author} {\bibfnamefont {Takahiro}\
  \bibnamefont {Morimoto}}, \bibinfo {author} {\bibfnamefont {Hoi~Chun}\
  \bibnamefont {Po}}, \ and\ \bibinfo {author} {\bibfnamefont {Ashvin}\
  \bibnamefont {Vishwanath}},\ }\bibfield  {title} {\enquote {\bibinfo {title}
  {{Floquet topological phases protected by time glide symmetry}},}\ }\href
  {\doibase 10.1103/PHYSREVB.95.195155/FIGURES/5/MEDIUM} {\bibfield  {journal}
  {\bibinfo  {journal} {Physical Review B}\ }\textbf {\bibinfo {volume} {95}},\
  \bibinfo {pages} {195155} (\bibinfo {year} {2017})}\BibitemShut {NoStop}%
\bibitem [{\citenamefont {Peng}\ and\ \citenamefont {Refael}(2019)}]{Peng2019}%
  \BibitemOpen
  \bibfield  {author} {\bibinfo {author} {\bibfnamefont {Yang}\ \bibnamefont
  {Peng}}\ and\ \bibinfo {author} {\bibfnamefont {Gil}\ \bibnamefont
  {Refael}},\ }\bibfield  {title} {\enquote {\bibinfo {title} {Floquet
  second-order topological insulators from nonsymmorphic space-time
  symmetries},}\ }\href {\doibase 10.1103/PhysRevLett.123.016806} {\bibfield
  {journal} {\bibinfo  {journal} {Phys. Rev. Lett.}\ }\textbf {\bibinfo
  {volume} {123}},\ \bibinfo {pages} {016806} (\bibinfo {year}
  {2019})}\BibitemShut {NoStop}%
\bibitem [{\citenamefont {Peng}(2020)}]{Peng2020}%
  \BibitemOpen
  \bibfield  {author} {\bibinfo {author} {\bibfnamefont {Yang}\ \bibnamefont
  {Peng}},\ }\bibfield  {title} {\enquote {\bibinfo {title} {Floquet
  higher-order topological insulators and superconductors with space-time
  symmetries},}\ }\href {\doibase 10.1103/PhysRevResearch.2.013124} {\bibfield
  {journal} {\bibinfo  {journal} {Phys. Rev. Res.}\ }\textbf {\bibinfo {volume}
  {2}},\ \bibinfo {pages} {013124} (\bibinfo {year} {2020})}\BibitemShut
  {NoStop}%
\bibitem [{\citenamefont {Peng}\ \emph {et~al.}(2021)\citenamefont {Peng},
  \citenamefont {Haim}, \citenamefont {Karzig}, \citenamefont {Peng},\ and\
  \citenamefont {Refael}}]{Peng2021}%
  \BibitemOpen
  \bibfield  {author} {\bibinfo {author} {\bibfnamefont {Changnan}\
  \bibnamefont {Peng}}, \bibinfo {author} {\bibfnamefont {Arbel}\ \bibnamefont
  {Haim}}, \bibinfo {author} {\bibfnamefont {Torsten}\ \bibnamefont {Karzig}},
  \bibinfo {author} {\bibfnamefont {Yang}\ \bibnamefont {Peng}}, \ and\
  \bibinfo {author} {\bibfnamefont {Gil}\ \bibnamefont {Refael}},\ }\bibfield
  {title} {\enquote {\bibinfo {title} {Floquet majorana bound states in
  voltage-biased planar josephson junctions},}\ }\href {\doibase
  10.1103/PhysRevResearch.3.023108} {\bibfield  {journal} {\bibinfo  {journal}
  {Phys. Rev. Research}\ }\textbf {\bibinfo {volume} {3}},\ \bibinfo {pages}
  {023108} (\bibinfo {year} {2021})}\BibitemShut {NoStop}%
\bibitem [{\citenamefont {Peng}(2022)}]{Peng2022}%
  \BibitemOpen
  \bibfield  {author} {\bibinfo {author} {\bibfnamefont {Yang}\ \bibnamefont
  {Peng}},\ }\bibfield  {title} {\enquote {\bibinfo {title} {Topological
  space-time crystal},}\ }\href {\doibase 10.1103/PhysRevLett.128.186802}
  {\bibfield  {journal} {\bibinfo  {journal} {Phys. Rev. Lett.}\ }\textbf
  {\bibinfo {volume} {128}},\ \bibinfo {pages} {186802} (\bibinfo {year}
  {2022})}\BibitemShut {NoStop}%
\bibitem [{\citenamefont {Katan}\ and\ \citenamefont
  {Podolsky}(2013)}]{Katan2013}%
  \BibitemOpen
  \bibfield  {author} {\bibinfo {author} {\bibfnamefont {Yaniv~Tenenbaum}\
  \bibnamefont {Katan}}\ and\ \bibinfo {author} {\bibfnamefont {Daniel}\
  \bibnamefont {Podolsky}},\ }\bibfield  {title} {\enquote {\bibinfo {title}
  {Modulated floquet topological insulators},}\ }\href {\doibase
  10.1103/PhysRevLett.110.016802} {\bibfield  {journal} {\bibinfo  {journal}
  {Phys. Rev. Lett.}\ }\textbf {\bibinfo {volume} {110}},\ \bibinfo {pages}
  {016802} (\bibinfo {year} {2013})}\BibitemShut {NoStop}%
\bibitem [{\citenamefont {Yao}\ \emph {et~al.}(2017)\citenamefont {Yao},
  \citenamefont {Yan},\ and\ \citenamefont {Wang}}]{Yao2017}%
  \BibitemOpen
  \bibfield  {author} {\bibinfo {author} {\bibfnamefont {Shunyu}\ \bibnamefont
  {Yao}}, \bibinfo {author} {\bibfnamefont {Zhongbo}\ \bibnamefont {Yan}}, \
  and\ \bibinfo {author} {\bibfnamefont {Zhong}\ \bibnamefont {Wang}},\
  }\bibfield  {title} {\enquote {\bibinfo {title} {Topological invariants of
  floquet systems: General formulation, special properties, and floquet
  topological defects},}\ }\href {\doibase 10.1103/PhysRevB.96.195303}
  {\bibfield  {journal} {\bibinfo  {journal} {Phys. Rev. B}\ }\textbf {\bibinfo
  {volume} {96}},\ \bibinfo {pages} {195303} (\bibinfo {year}
  {2017})}\BibitemShut {NoStop}%
\bibitem [{\citenamefont {Altland}\ and\ \citenamefont
  {Zirnbauer}(1997)}]{Altland1997}%
  \BibitemOpen
  \bibfield  {author} {\bibinfo {author} {\bibfnamefont {Alexander}\
  \bibnamefont {Altland}}\ and\ \bibinfo {author} {\bibfnamefont {Martin~R.}\
  \bibnamefont {Zirnbauer}},\ }\bibfield  {title} {\enquote {\bibinfo {title}
  {Nonstandard symmetry classes in mesoscopic normal-superconducting hybrid
  structures},}\ }\href {\doibase 10.1103/PhysRevB.55.1142} {\bibfield
  {journal} {\bibinfo  {journal} {Phys. Rev. B}\ }\textbf {\bibinfo {volume}
  {55}},\ \bibinfo {pages} {1142--1161} (\bibinfo {year} {1997})}\BibitemShut
  {NoStop}%
\bibitem [{\citenamefont {Po}\ \emph {et~al.}(2017)\citenamefont {Po},
  \citenamefont {Fidkowski}, \citenamefont {Vishwanath},\ and\ \citenamefont
  {Potter}}]{Po2017}%
  \BibitemOpen
  \bibfield  {author} {\bibinfo {author} {\bibfnamefont {Hoi~Chun}\
  \bibnamefont {Po}}, \bibinfo {author} {\bibfnamefont {Lukasz}\ \bibnamefont
  {Fidkowski}}, \bibinfo {author} {\bibfnamefont {Ashvin}\ \bibnamefont
  {Vishwanath}}, \ and\ \bibinfo {author} {\bibfnamefont {Andrew~C.}\
  \bibnamefont {Potter}},\ }\bibfield  {title} {\enquote {\bibinfo {title}
  {Radical chiral floquet phases in a periodically driven kitaev model and
  beyond},}\ }\href {\doibase 10.1103/PhysRevB.96.245116} {\bibfield  {journal}
  {\bibinfo  {journal} {Phys. Rev. B}\ }\textbf {\bibinfo {volume} {96}},\
  \bibinfo {pages} {245116} (\bibinfo {year} {2017})}\BibitemShut {NoStop}%
\bibitem [{\citenamefont {Fulga}\ \emph {et~al.}(2019)\citenamefont {Fulga},
  \citenamefont {Maksymenko}, \citenamefont {Rieder}, \citenamefont {Lindner},\
  and\ \citenamefont {Berg}}]{Fulga2019}%
  \BibitemOpen
  \bibfield  {author} {\bibinfo {author} {\bibfnamefont {I.~C.}\ \bibnamefont
  {Fulga}}, \bibinfo {author} {\bibfnamefont {M.}~\bibnamefont {Maksymenko}},
  \bibinfo {author} {\bibfnamefont {M.~T.}\ \bibnamefont {Rieder}}, \bibinfo
  {author} {\bibfnamefont {N.~H.}\ \bibnamefont {Lindner}}, \ and\ \bibinfo
  {author} {\bibfnamefont {E.}~\bibnamefont {Berg}},\ }\bibfield  {title}
  {\enquote {\bibinfo {title} {{Topology and localization of a periodically
  driven Kitaev model}},}\ }\href
  {https://journals.aps.org/prb/abstract/10.1103/PhysRevB.99.235408} {\bibfield
   {journal} {\bibinfo  {journal} {Physical Review B}\ }\textbf {\bibinfo
  {volume} {99}},\ \bibinfo {pages} {235408} (\bibinfo {year}
  {2019})}\BibitemShut {NoStop}%
\bibitem [{\citenamefont {Kitaev}(2006)}]{Kitaev2006}%
  \BibitemOpen
  \bibfield  {author} {\bibinfo {author} {\bibfnamefont {Alexei}\ \bibnamefont
  {Kitaev}},\ }\bibfield  {title} {\enquote {\bibinfo {title} {Anyons in an
  exactly solved model and beyond},}\ }\href
  {https://www.sciencedirect.com/science/article/abs/pii/S0003491605002381}
  {\bibfield  {journal} {\bibinfo  {journal} {Annals of Physics}\ }\textbf
  {\bibinfo {volume} {321}},\ \bibinfo {pages} {2--111} (\bibinfo {year}
  {2006})}\BibitemShut {NoStop}%
\bibitem [{Note1()}]{Note1}%
  \BibitemOpen
  \bibinfo {note} {By time reversal symmetry, we mean that there is an
  anti-unitary operator $\protect \mathcal {T}$ such that the time-dependent
  Hamiltonian satisfies $H(t)=\protect \mathcal {T} H(T-t) \protect \mathcal
  {T}^{-1}$, for an appropriate choice of the origin of time.}\BibitemShut
  {Stop}%
\bibitem [{Note2()}]{Note2}%
  \BibitemOpen
  \bibinfo {note} {In the presence of the time vortex, the Hamiltonian is not
  translationally invariant, but has a combined space-time translation
  symmetry. The label $k$ of $\chi _{nk}$ is the eigenvalue of the Floquet
  eigenstate under this combined space-time translation; see Sec.~\ref {app:
  floquet eigenstates with tv}.}\BibitemShut {Stop}%
\bibitem [{Note3()}]{Note3}%
  \BibitemOpen
  \bibinfo {note} {In particular, if $k_0=0$, the system with $n_v=0$ has a
  $\pi $ mode at the edges and the $n_v=1$ system does not. If $k_0=\pi /a_0$,
  the $n_v=0$ system has a $\pi $ mode if $L_x/a_0$ (the number of sites around
  the cylinder) is even, and does not have one if $L_x/a_0$ is odd. In the
  $n_v=1$ case, the situation is reversed.}\BibitemShut {Stop}%
\bibitem [{\citenamefont {Uhrig}\ \emph {et~al.}(2019)\citenamefont {Uhrig},
  \citenamefont {Kalthoff},\ and\ \citenamefont {Freericks}}]{Uhrig_2019}%
  \BibitemOpen
  \bibfield  {author} {\bibinfo {author} {\bibfnamefont {Götz~S.}\
  \bibnamefont {Uhrig}}, \bibinfo {author} {\bibfnamefont {Mona~H.}\
  \bibnamefont {Kalthoff}}, \ and\ \bibinfo {author} {\bibfnamefont {James~K.}\
  \bibnamefont {Freericks}},\ }\bibfield  {title} {\enquote {\bibinfo {title}
  {Positivity of the spectral densities of retarded floquet green functions},}\
  }\href {\doibase 10.1103/physrevlett.122.130604} {\bibfield  {journal}
  {\bibinfo  {journal} {Physical Review Letters}\ }\textbf {\bibinfo {volume}
  {122}} (\bibinfo {year} {2019}),\ 10.1103/physrevlett.122.130604}\BibitemShut
  {NoStop}%
\bibitem [{\citenamefont {Teo}\ and\ \citenamefont {Kane}(2010)}]{Teo2010}%
  \BibitemOpen
  \bibfield  {author} {\bibinfo {author} {\bibfnamefont {Jeffrey~C.Y.}\
  \bibnamefont {Teo}}\ and\ \bibinfo {author} {\bibfnamefont {C.~L.}\
  \bibnamefont {Kane}},\ }\bibfield  {title} {\enquote {\bibinfo {title}
  {{Topological defects and gapless modes in insulators and
  superconductors}},}\ }\href {\doibase
  10.1103/PHYSREVB.82.115120/FIGURES/15/MEDIUM} {\bibfield  {journal} {\bibinfo
   {journal} {Physical Review B - Condensed Matter and Materials Physics}\
  }\textbf {\bibinfo {volume} {82}},\ \bibinfo {pages} {115120} (\bibinfo
  {year} {2010})}\BibitemShut {NoStop}%
\bibitem [{\citenamefont {Kitaev}(2001)}]{Kitaev2001}%
  \BibitemOpen
  \bibfield  {author} {\bibinfo {author} {\bibfnamefont {A.~Yu}\ \bibnamefont
  {Kitaev}},\ }\bibfield  {title} {\enquote {\bibinfo {title} {{Unpaired
  Majorana fermions in quantum wires}},}\ }\href {\doibase
  10.1070/1063-7869/44/10S/S29} {\bibfield  {journal} {\bibinfo  {journal}
  {Phys. Usp.}\ }\textbf {\bibinfo {volume} {44}},\ \bibinfo {pages} {131--136}
  (\bibinfo {year} {2001})},\ \Eprint {http://arxiv.org/abs/cond-mat/0010440}
  {arXiv:cond-mat/0010440} \BibitemShut {NoStop}%
\bibitem [{\citenamefont {Keselman}\ \emph {et~al.}(2013)\citenamefont
  {Keselman}, \citenamefont {Fu}, \citenamefont {Stern},\ and\ \citenamefont
  {Berg}}]{Keselman2013}%
  \BibitemOpen
  \bibfield  {author} {\bibinfo {author} {\bibfnamefont {Anna}\ \bibnamefont
  {Keselman}}, \bibinfo {author} {\bibfnamefont {Liang}\ \bibnamefont {Fu}},
  \bibinfo {author} {\bibfnamefont {Ady}\ \bibnamefont {Stern}}, \ and\
  \bibinfo {author} {\bibfnamefont {Erez}\ \bibnamefont {Berg}},\ }\bibfield
  {title} {\enquote {\bibinfo {title} {{Inducing time-reversal-invariant
  topological superconductivity and fermion parity pumping in quantum
  wires}},}\ }\href {\doibase 10.1103/PHYSREVLETT.111.116402/FIGURES/3/MEDIUM}
  {\bibfield  {journal} {\bibinfo  {journal} {Physical Review Letters}\
  }\textbf {\bibinfo {volume} {111}},\ \bibinfo {pages} {116402} (\bibinfo
  {year} {2013})}\BibitemShut {NoStop}%
\bibitem [{Note4()}]{Note4}%
  \BibitemOpen
  \bibinfo {note} {If $\nu (\varepsilon , k_x)=1$, then the gap at the end of
  the system has to close at quasi-energy $\varepsilon $ for some value of
  $\lambda $. This, in turn, implies a topological gap closing along the ray at
  the same $k_x$ at this value of $\lambda $ (i.e., a level crossing of two
  states related by particle-hole symmetry). Viewed as an adiabatic process,
  such a level crossing changes the fermion parity corresponding to the Dirac
  fermion formed by the two Majorana states that cross, hence it is referred to
  as a fermion parity pump.}\BibitemShut {Stop}%
\bibitem [{Note5()}]{Note5}%
  \BibitemOpen
  \bibinfo {note} {Since $H_{\protect \rm {eff}}(\protect \bm {k},t;2\pi
  \lambda )$ is assumed to be gapped at $\varepsilon =\protect \tfrac {\pi
  }{T}$ for this value of $t$, we can use the usual convention for the branch
  cuts of $\log (z)$, with a discontinuity along $\protect \rm {Re(z)<0}$,
  $\protect \rm {Im}(z)=0$}\BibitemShut {NoStop}%
\bibitem [{lin()}]{link_interactive}%
  \BibitemOpen
  \href@noop {} {}\bibinfo {note} {An interactive version of the
  three-dimensional figures can be downloaded from
  \href{https://github.com/kishonyWIS/free_fermions/tree/FloquetKSL/graphs/time_vortex}{this
  link}.}\BibitemShut {Stop}%
\end{thebibliography}%

\end{document}